\documentclass[journal=jacsat,manuscript=article]{achemso}
\usepackage[version=3]{mhchem} 
\usepackage[Euler]{upgreek}
\usepackage{mathptmx}
\usepackage{etoolbox}
\usepackage[mathcal,mathscr]{eucal}
\usepackage{times}
\usepackage{lmodern}
\usepackage{enumerate}
\usepackage{textcomp}
\usepackage{stmaryrd}
\usepackage[colorlinks=true, allcolors=black]{hyperref}
\usepackage{footmisc}
\usepackage{tikz,xcolor,hyperref}
\usepackage{footmisc}
\usepackage{tikz,xcolor,hyperref}
\newcommand{\orcid}[1]{\href{https://orcid.org/#1}{\includegraphics[width=8pt]{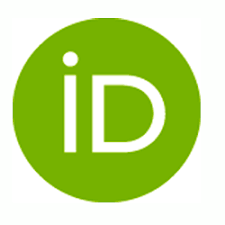}}}
\usepackage[numbers,super]{natbib}
\setkeys{acs}{maxauthors = 20}
\SectionNumbersOn

\definecolor{lime}{HTML}{A6CE39}

\newcommand{\onlinecite}[1]{\hspace{-1 ex} \nocite{#1}\citenum{#1}}

\author{Azimatu Fangnon}
\affiliation{Department of Applied Physics, Aalto University, FI-00076 AALTO, Finland}
\email{azimatu.fangnon@aalto.fi}
\author{Marc Dvorak}
\affiliation{Department of Applied Physics, Aalto University, FI-00076 AALTO, Finland}
\author{Ville Havu}
\affiliation{Department of Applied Physics, Aalto University, FI-00076 AALTO, Finland}
\author{Milica Todorovi{\'c}}
\affiliation{Department of Mechanical and Materials Engineering, University of Turku, FI-02458963 Turku, Finland}
\author{Jingrui Li}
\affiliation{Electronic Materials Research Laboratory, Key Laboratory of the Ministry of Education \& International Center for Dielectric Research, School of Electronic Science and Engineering, Xi'an Jiaotong University, Xi'an 710049, China}
\author{Patrick Rinke}
\affiliation{Department of Applied Physics, Aalto University, FI-00076 AALTO, Finland}

\title[Protective Coating Interfaces for Perovskite Solar Cell Materials: A
First Principles Study]{Protective Coating Interfaces for Perovskite Solar Cell Materials: A First Principles Study}

\abbreviations{IR,NMR,UV}

\begin{document}

\begin{tocentry}

\centering
\includegraphics[width=\textwidth, keepaspectratio]{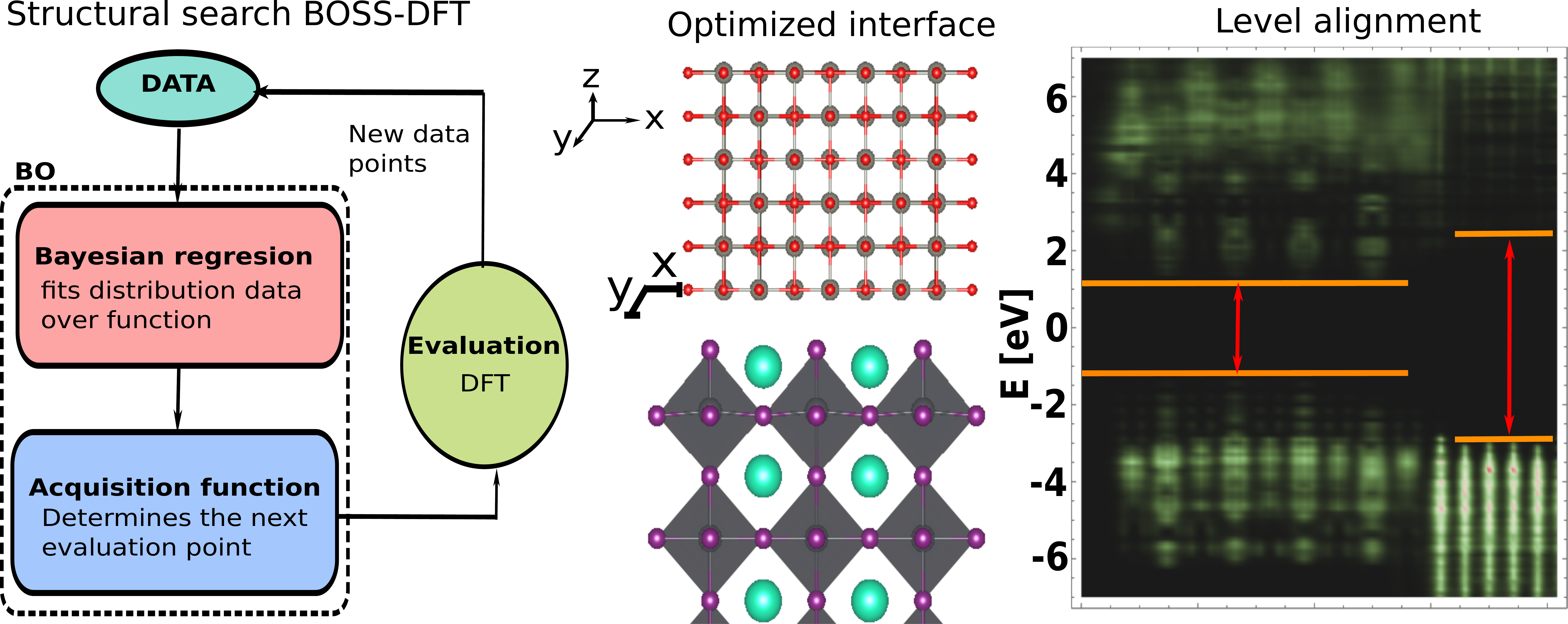}

\end{tocentry}

\begin{abstract}
  The protection of halide perovskites is important for the performance and stability of emergent perovskite-based optoelectronic technologies. In this work, we investigate the potential inorganic protective coating materials ZnO, SrZrO$_3$\,, and ZrO$_2$ for the CsPbI$_3$ perovskite. The optimal interface registries are identified with Bayesian optimization. We then use semi-local density-functional theory (DFT) to determine the atomic structure at the interfaces of each coating material with the clean CsI-terminated surface and three  reconstructed surface models with added PbI$_2$ and CsI complexes. For the final structures, we explore the level alignment at the interface with hybrid DFT calculations. Our analysis of the level alignment at the coating-substrate interfaces reveals no detrimental mid-gap states, but substrate-dependent valence and conduction band offsets. While ZnO and SrZrO$_3$ act as insulators on CsPbI$_3$, ZrO$_2$ might be suitable as electron transport layer with the right interface engineering.
\end{abstract}

\section*{Keywords}
Interface, Surface, level alignment, Perovskite, Transport layer, Coating, Density functional theory, Bayesian optimization.

\section{Introduction}
\label{intro}

Halide perovskites (HPs) have emerged as 
promising materials for next-generation optoelectronics, as evidenced by the steep  rise in power conversion efficiency of perovskite solar cells (PSCs) from 10\% \cite{Kim2012,Lee2012} to 25.5\% \cite{MinH2021} within one decade. Other viable HP applications are light-emitting diodes, lasers, and photodetectors. \cite{ChenP2018,ZhangC2020,DongH2020,WangY2021} HPs and HP-based devices are particularly attractive due to their ease of fabrication, low processing temperature, cost effectiveness, and availability of raw materials. \cite{CorreaBaena2017,Rajagopal2018,LiH2020} 
Despite HPs' exceptional optoelectronic properties, their large-scale production and commercialization is still impeded by several factors: The commonly-used hybrid (organic-inorganic) HPs are known to suffer from rapid degradation when exposed to moisture, heat, or oxygen. \cite{NiuG14,NiuG15,HuangJ17,KimGH17,Mesquita18,LiF18} 
In addition, organic charge-transport-layer materials will generally limit the device performance due to their structural and chemical disorder. \citep*{Shao2015} For example, \textit{Spiro}-OMeTAD, which is the most common hole-transport-layer (HTL) material in PSCs \cite{Kim2012, Lee2012, Leijtens2013, Im2011,Burschka2013}, suffers from instability, low hole mobility and conductivity \cite{Burschka2011}, and undesirable impact on PSC stability \cite{Saliba16b}. Surface passivation with more stable materials \cite{SchmidtL14, SoranyelG15,Dong19}, especially inorganic ones, is thus important to mitigate the negative effects of ambient conditions on HP materials and devices.

These challenges could be addressed with an all-inorganic strategy. \cite{ChenW2020,LiuC2020,ZhangL2020} In this strategy, inorganic materials are chosen for all device layers:  (mostly Cs-based) perovskites for the photo-absorbing or emitting layer, inorganic semiconductors as electron- and hole-transport layers, and inorganic insulators as charge-blocking materials. In this context, protective inorganic coatings have been proposed. \cite{Matteocci16,Cheacharoen18a,Cheacharoen18b,Seidu19} It would be particularly beneficial, if the coating materials could also serve as electron-transport (ETL) or as hole-transport layers (HTL). So far, the studied inorganic interlayer materials in perovskite-based optoelectronic devices are mainly common binary compounds. Typical examples include TiO$_2$ \cite{Lian2018, Tong2016, Wang2019}, SnO$_2$ \cite{Anaraki2016,Jiang18, Xiong2018, Lin2021}, and ZnO \cite{Lin2021, Tong2016} for ETLs, CuI \cite{Christians2014, Huangfu2015}, NiO \cite{Kim2015,Islam2017, Chen2015, Subbiah2014}, graphene oxide \cite{Wu2014}, and CuSCN \cite{Subbiah2014} for HTLs, as well as alkali-metal halides for charge-blocking layers \cite{ShiY2018,XuL2020,YuanF2020,XuJ2021}.

\begin{figure}[!htp]
\centering
\includegraphics[width=\linewidth]{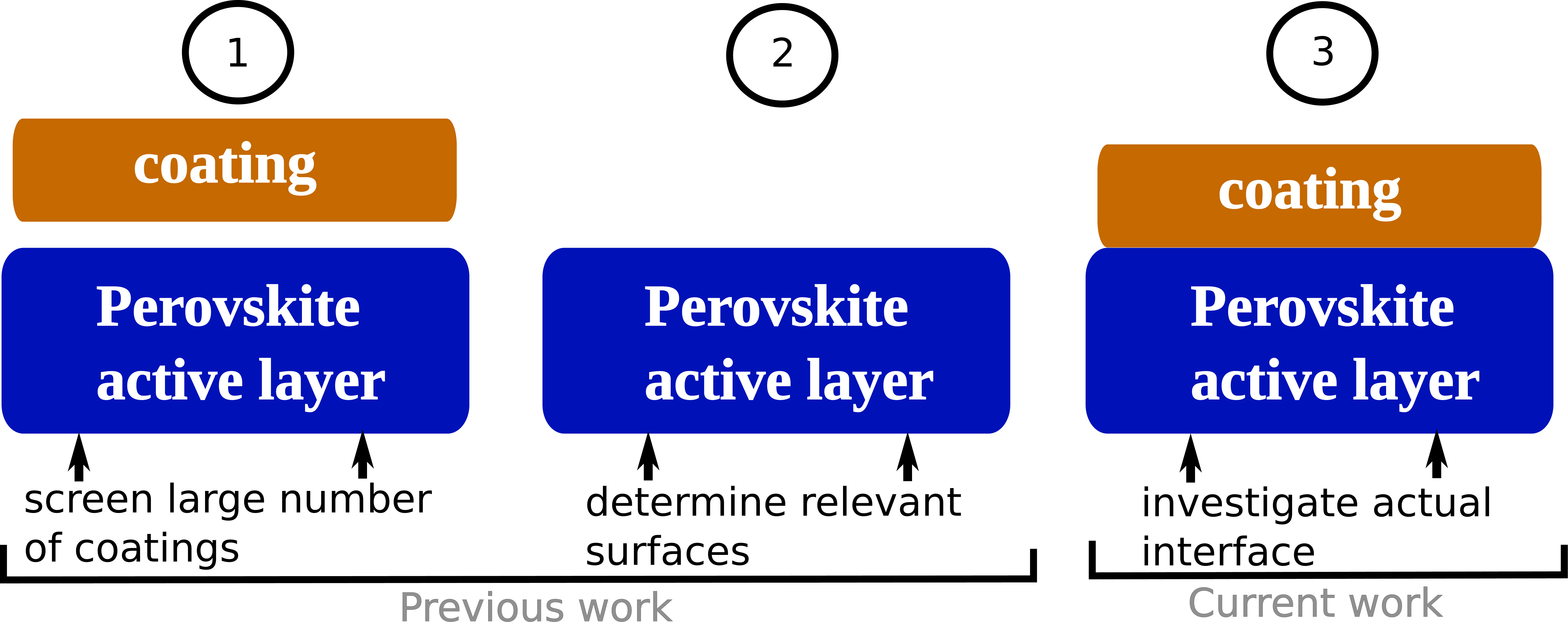}\vspace{0.1em}
\caption{Conceptual workflow of identifying and characterizing suitable perovskite coating materials.}
\label{schematics}
\end{figure}

In our previous work, we applied a data-driven approach to discover inorganic materials suitable for perovskite-based devices with the aim to further enhance the device performance and stability \cite{Seidu19}. We developed a three-stage scheme as shown in Fig.~\ref{schematics}. At stage 1, we screened a materials database for inorganic coating materials that meet a series of requirements such as band gap, stability, transport properties, and crystal structure.\cite{Seidu19} Stage 2 identifies the stable surface structures of $\text{CsPbI}_3^{}$ and $\text{CH}_3^{}\text{NH}_3^{}\text{PbI}_3^{}$ at different growth conditions, for which we carried out surface-phase-diagram analysis based on density-functional-theory (DFT) and \textit{ab initio} thermodynamics.\cite{Seidu2021,ASeidu2021}  

This work presents stage 3 that computationally estimates whether the materials which pass the screening at stage 1 \cite{Seidu19} are indeed good coating candidates for HPs. To this end, we investigate the interfaces between the candidate coatings and a series of HP surface models produced at stage 2 \cite{Seidu2021} using DFT. The surface registry match between the perovskite and the coating was determined by Bayesian optimization (BO) active learning, based on the minima in the total energy landscape. The minima of the energy landscape corresponds to the binding energy (E$_\text{b}$) of the most stable registry. 
We choose $\upalpha$-$\text{CsPbI}_3^{}$ as the prototype HP model system for this first case study, while similar systems such as $\upgamma$-$\text{CsPbI}_3^{}$ and $\text{CH}_3^{}\text{NH}_3^{}\text{PbI}_3^{}$ will be the subject of future work. 

The remainder of this paper is organized as follows: We present a brief description of our computational approach in Section~\ref{comp}. In Section~\ref{results}, we present the results from the BOSS/DFT calculations, analyze the features of the optimized interface structures, and establish the level alignments at the coating-perovskite interface for all coatings and substrates. Section~\ref{discussion} presents a brief discussion of our results. We conclude with a summary of our findings in Section~\ref{conclusion}. Computational details are presented in Section~\ref{method}.

\section{Computational approach}
\label{comp}
\begin{figure}[!ht]
\centering
\includegraphics[width=\linewidth]{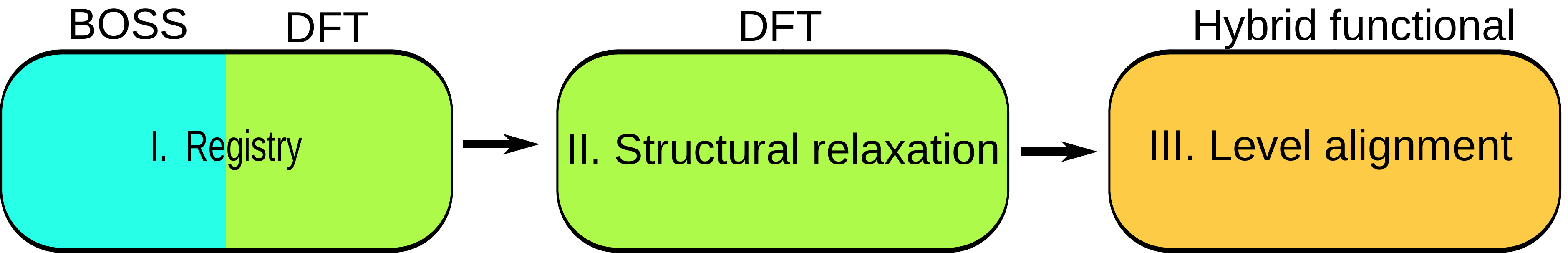}\vspace{0.1em}
\caption{Protocol for computational characterization of coating-perovskite interfaces with BOSS and DFT.} 
\label{boss}
\end{figure}

We designed a three-step protocol for our interface study as sketched in Fig.~\ref{boss}. Step I employs the recently developed Bayesian Optimisation Structure Search (BOSS) package \cite{Todorovic2019} to determine the registry between the coating and the HP substrate. Starting from this registry, the atomic positions of the coating-perovskite interface are relaxed with DFT in step II. In step III, interfacial electronic structure calculations are carried out for the relaxed interfaces using hybrid functional DFT.

We propose such a two-step (steps I and II) structure-search strategy based on the following consideration. The most stable interface geometry corresponds to the global minimum of the total-energy landscape of the coating-perovskite combination, which is a multi-dimensional function of a series of parameters that define the relative geometry between these two components. This total-energy landscape is  very complex due to the poly-atomic nature of both coating and HP surfaces. Therefore it is not easy to identify the most stable structure with simple DFT structure relaxation, as different initial structures might fall to different local minima and miss the global minimum. 

BOSS has already shown its power in problems such as conformer search for organic molecules\cite{Todorovic2019,FangL2021}, organic molecule adsorption at semiconductor surfaces \cite{Todorovic2019}, and film growth of organic adsorbates on metallic surfaces \cite{EggerAT2020,Jarvi2020,Jarvi2021}. In this work, we employ BOSS to tackle the interface problem between two inorganic materials (coating and HP) in step I, which we believe can efficiently narrow down the search space for further geometry relaxation in step II. In our BOSS search, state-of-the-art single-point DFT calculations are performed for different structures, and BOSS correlates the structures with an energy landscape through active learning with a BO algorithm. A surrogate model is fitted to the DFT data points employing Gaussian process regression (GPR), which is refined by acquiring further data with a smart sampling strategy. In such a way, a relatively modest number of DFT data points suffice to converge the multi-dimensional total-energy landscape. 

For a semi-infinite slab with a thin coating layer, the band gap deep into the bulk is the same for any surface reconstruction or defect. For this reason and to provide a consistent ranking of level alignments among coatings, we use the bulk as a  ``model substrate'' for all the interfaces. For each interface, we extracted the valence (VB) and conduction band (CB) offsets from the spatially resolved local density of states (LDOS) of Fig.~S5 in SI. The offsets are then added (or subtracted) from the bulk conduction band minimum (CBM) (or valence band maximum, VBM) to create the alignments here. Our bulk band structure and LDOS plots are based on a hybrid functional. We also included spin-orbit coupling (SOC) to the bulk band structure calculation. Base on our previous works, we expect the inclusion of SOC to shift the CBM down (into the gap), which will lead to a reduction of the band gap energy. To account for the changes resulting from SOC, we will shift the VBM and CBM of the bulk LDOS for our final coating-perovskite interface level alignments. Details of these calculations are outlined in Section~\ref{method}.

\section{Results}\label{results}

In this work, we find the stable registry and interface structure, analyse the electronic properties, and establish the level alignment at the coating-perovskite interfaces. We considered SrZrO$_3$ (cubic, $Pm\bar{3}m$), ZnO (cubic, $Fm\bar{3}m$) and ZrO$_2$ (tetragonal $P4_2/nmc$) as coatings based on results from our previous study \cite{Seidu19}.  For the substrate, we investigated both the ideal clean CsI-terminated (CsI-T) and reconstructed surface models with adatoms (of CsPbI$_3$) to simulate different synthesis conditions. The selected surface reconstructions are taken from Ref.~\onlinecite{Seidu2021}, in which they were determined to be the most relevant reconstructions. We used the stable reconstructed surface models clean CsI-T surface, $i_{\text{PbI}_2^{}}$, $i_{2\text{PbI}_2^{}}$ and $i_{4\text{CsI}}$ of pm\=3m CsPbI$_3$. $i_X$ denotes adatom structures with $X$ = PbI$_2$, 2PbI$_2$ or 4CsI. Details of the computations are outlined in Section~\ref{method}. This section will present and discuss the binding energy landscapes and optimized interface structures for ZnO on the four substrates as a prototype. We use ZnO as a prototype because, of all the coatings in this work, it is the most studied transport layer in PSCs. Similar results for the other coatings are presented in the Supplementary Information (SI). The level alignments at the coating/perovskite interface for all the structures in this work will  also be presented and discussed.

\subsection{Binding energy landscapes from BOSS}

\begin{figure}[!ht]
\centering
\includegraphics[width=\linewidth]{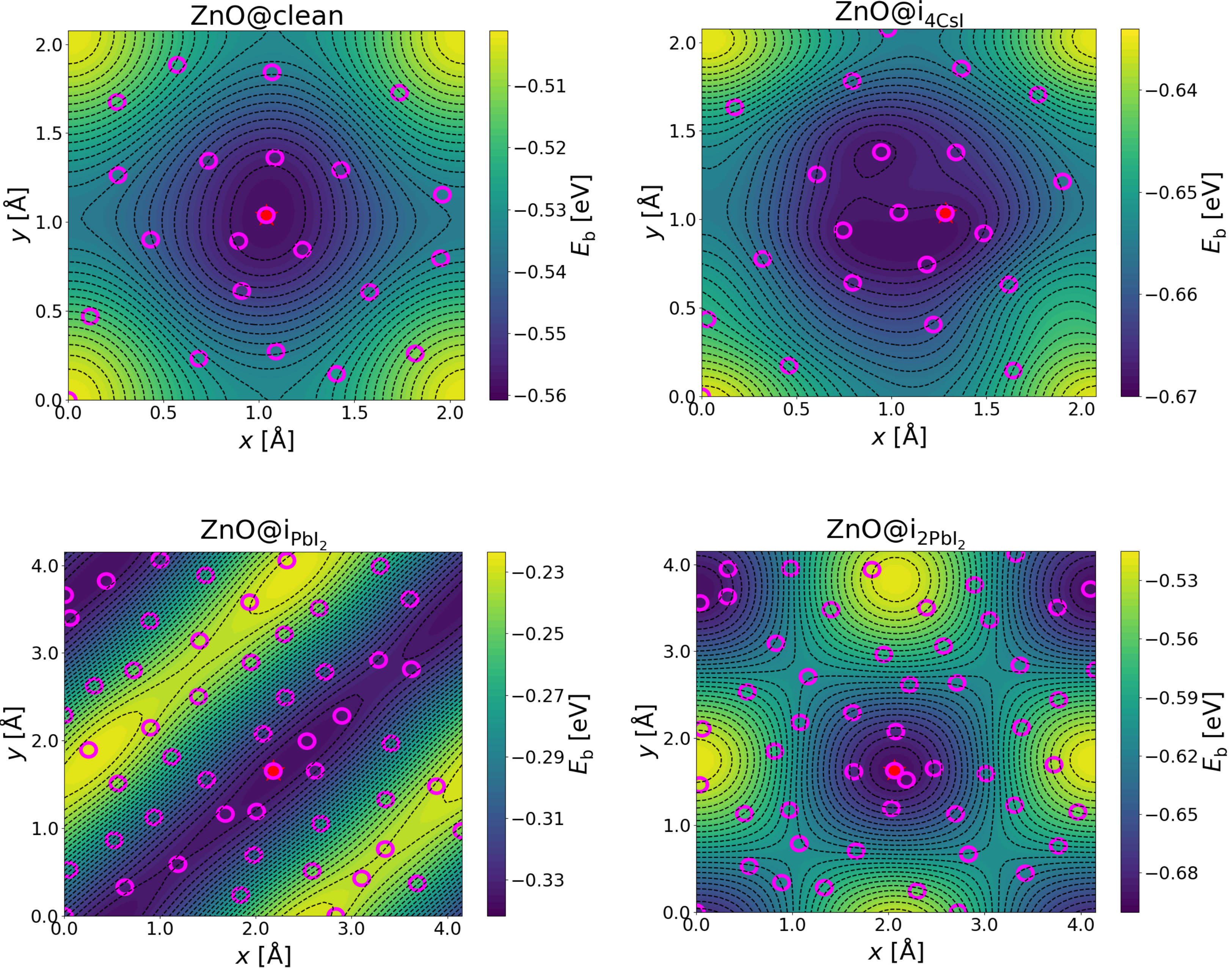}\vspace{0.1em}
\caption {Binding energy landscapes of perovskite/coating with BOSS. The pink circles and red star depict the acquisition points and minimum potential energies, respectively, as the coating is translated by $x$, $y$ on the substrate. Where $x$, $y$ describe the confines of the periodic unit on the surface.} \label{landscape}
\end{figure}

Figure~\ref{landscape} shows the 2-dimensional (2D) binding energy landscapes for ZnO@clean, ZnO@$i_{\text{PbI}_2^{}}$, ZnO@$i_{\text{2PbI}_2^{}}$ and ZnO@$i_{4\text{CsI}}$. The 2D binding energy landscapes for SrZrO$_3$ and ZrO$_2$ on the four substrates are shown in Fig.~S1 of the SI. The pink circles depict the acquisition points and the red star the energy minimum, which corresponds to the minimum binding energy (E$_b$) of the surrogate models. The yellow shades in the color map depict regions of high binding energy followed by the green shades with the dark blue shades showing the lowest  energies. Here, $x$ and $y$ are the translations of the coating in the $x$ and $y$ directions from some initial origin. Due to the periodicity of the system in $x-y$, these displacements describe the entire search domain. 

The binding energy landscapes for ZnO@clean and ZnO@$i_{4\text{CsI}}$ (top panel) show an $x$ and $y$ range of $0.00-2.08$ {\AA}, representing the smallest unit cell of the coating@perovskite interface. For a $3\times3\times3$ coating on a $2\times2$ substrate, the primitive unit cell of the ensemble is 1/6 of the supercell, which corresponds to a search space of $0.00-2.08$ {\AA}. In the case of ZnO@$i_{\text{PbI}_2^{}}$ and ZnO@$i_{2\text{PbI}_2^{}}$, the interface commensurability changes due to the added PbI$_2$ complexes. For these, we used a search space of 1/3 the supercell corresponding to $0.00-4.15$ {\AA}. 
The registries with the lowest binding energies are listed in Tab.~S2 of the SI.

Due to the similar search space for ZnO@clean and ZnO@$i_{4\text{CsI}}$, the binding energy landscapes are almost the same and their minimum energies (red star) are both located at the center of the landscape. In ZnO@$i_{\text{PbI}_2}$, the landscape looks different due to the broken periodicity emanating from the added PbI$_2$ unit. Similarly, the pattern in ZnO@$i_{2\text{PbI}_2}$ is different. Here, a diagonal periodicity is seen due to the repeated PbI$_2$ unit on the surface of the substrate.

\subsection{Optimized interface structures}
\begin{figure}[!ht]
\centering
\includegraphics[width=\linewidth]{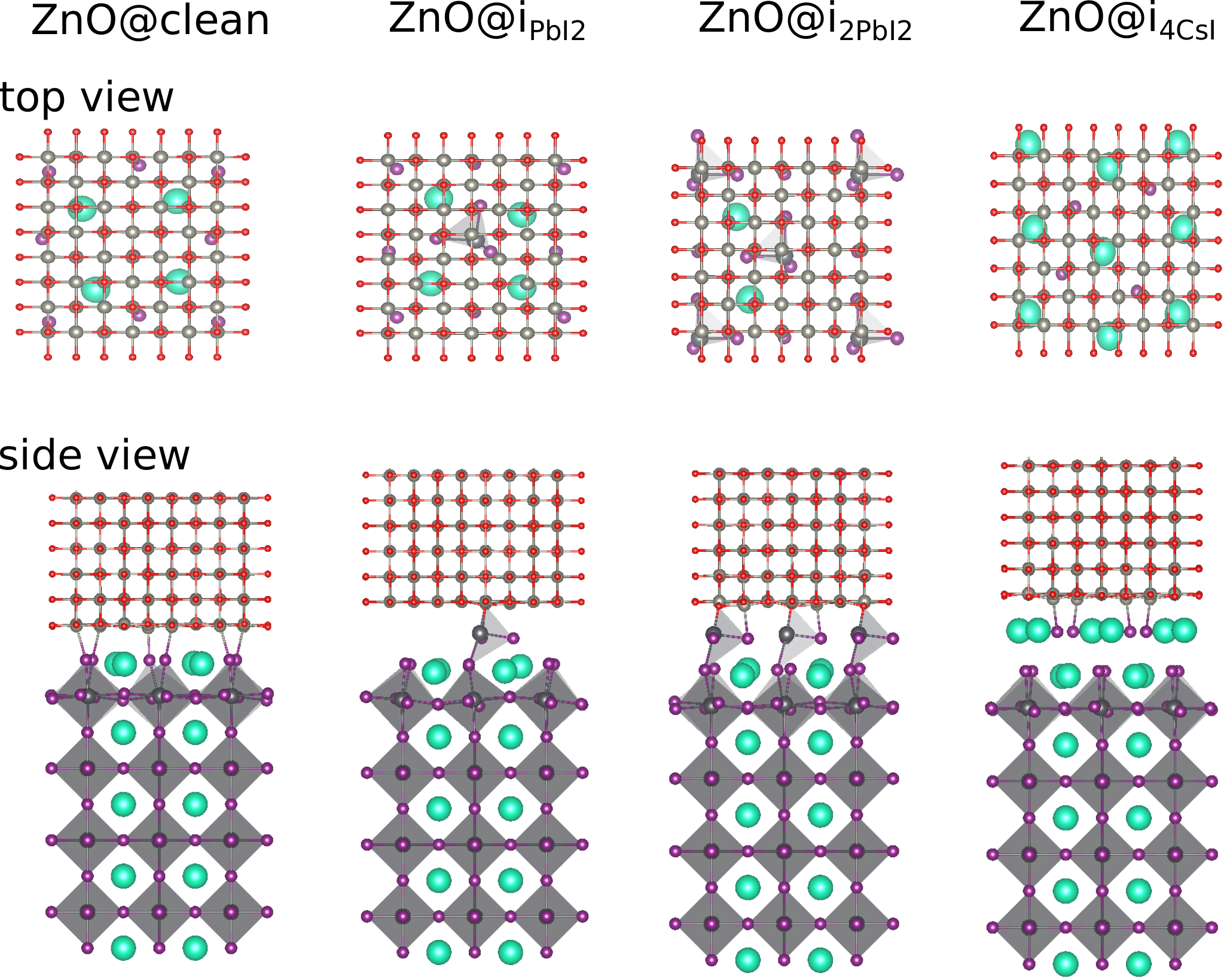}\vspace{0.1em}
\caption {Optimized structure of ZnO on the clean surface and its reconstructed models: From left to right are ZnO@clean,  ZnO@$i_{\text{PbI}_2^{}}$, ZnO@$i_{2\text{PbI}_2^{}}$ and  ZnO@$i_{4\text{CsI}}$. $\text{Cs}$, $\text{Pb}$, $\text{I}$, Zn and O are colored in green, black, purple, light grey and red, respectively. The PbI$_6$ octahedra are colored in dark gray.} 
\label{structure}
\end{figure}

Figure~\ref{structure} depicts the optimized interface structures obtained from DFT relaxations of the optimal registry positions for ZnO@clean, ZnO@$i_{\text{PbI}_2^{}}$, ZnO@$i_{2\text{PbI}_2^{}}$ and ZnO@$i_{4\text{CsI}}$. Similar results for SrZrO$_3$ and ZrO$_2$ are shown in Fig.~S2 of the SI. Detailed structural and computational information are outlined in Section~\ref{method}. We observe a rearrangement of atoms in the substrates to accommodate the lattice strain between the coatings and substrates. Specifically, the Cs-I bond lengths in the topmost substrate layers change slightly to accommodate the bonding between the bottom layers of the coatings and the CsI layers. This is more pronounced in ZnO@$i_{4\text{CsI}}$ where the topmost CsI layer is pulled into the coatings. Similarly, the Pb-I bonds in the topmost polyhedra of ZnO@$i_{\text{PbI}_2^{}}$ and ZnO@$i_{2\text{PbI}_2^{}}$ tilt slightly at the interface. We also observe similar features for ZrO$_2$- and SrZrO$_3$-based interfaces (Fig.~S2). Despite these changes in bond lengths, our coatings show strong bonding with the substrates with no visible structural distortions.

Table~\ref{bind_energy} shows the optimized binding energies and lattice strain for all three coatings on our four most relevant reconstructed $\upalpha$-CsPbI$_3$ surface models. Here, $E_b = E_{\text{ensemble}} - ( E_{\text{coating}} +  E_{\text{substrate}})$.  
The absolute lattice strain increases from SrZrO$_3$ to ZrO$_2$ with ZnO having the largest. Similarly, binding reduces with an absolute increase in lattice strains (i.e. the absolute value of the binding energy decreases). Specifically, substrates with SrZrO$_3$ coatings show the strongest binding followed by those with ZrO$_2$ and ZnO coatings (Table~\ref{bind_energy}). This observation can be simply explained by the fact that structures with larger mismatch have a greater energetic cost due to strain, hence leading to decrease in binding strength. Figure~S3 of the SI shows a plot of binding energies as a function of lattice strain. Structurally, we also observe minimum rearrangement in atomic positions for ZnO and ZrO$_2$ on all substrates. SrZrO$_3$ on the other hand, shows significant rearrangement causing shifts in atomic positions (Fig.~S2). These observations could also contribute to the varying binding energies as seen in Table~\ref{bind_energy}.

\begin{table}[!htp]
\caption{Binding energies (in eV) and lattice strain (in \%) of the three coatings on the four most relevant  $\upalpha$-CsPbI$_3$ reconstructed surface models.} \label{bind_energy}

\begin{tabular}{c@{\hspace{2em}}c@{\hspace{2em}} c@{\hspace{2em}} c@{\hspace{2em}} c@{\hspace{1em}}r} 
& Strain & \multicolumn{4}{c}{Binding Energy [eV]} 
\\ \hline 
\hline
& [\%] & ~clean & ~$i_{\text{PbI}_2^{}}$ & ~$i_{2\text{PbI}_2^{}}$ &~$i_{4\text{CsI}}$
\\\hline
SrZrO$_3$ & ~1.0 & $-5.53$ & $-5.38$ & $-7.03$ & $-8.16$ \\ 
ZrO$_2$  &~-2.2 & $-2.14$ & $-3.14$ & $-4.85$ & $-4.23$ \\
ZnO   & ~4.5    & $-1.15$ & $-0.73$ & $-1.32$ & $-2.11$ \\ 
 \hline \hline
\end{tabular}
\end{table}

\subsection{Level alignment of coating-perovskite interfaces}

\begin{figure}[!ht]
\centering
\includegraphics[width=\textwidth]{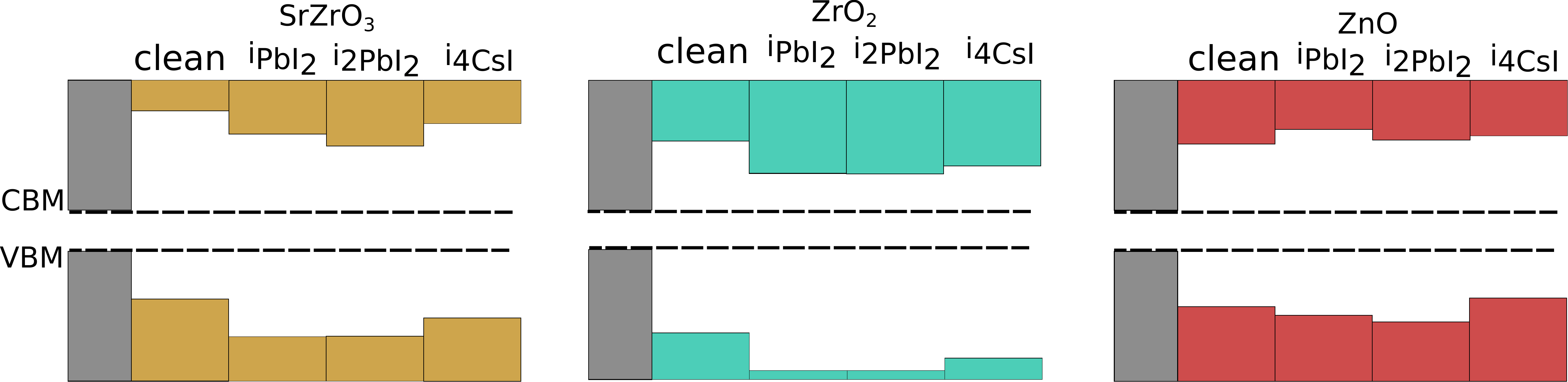}\vspace{0.1em}
\caption {Band alignment at the coating-perovskite interface. The grey, red, yellow and green shaded regions are representative of the bulk substrate, ZnO, SrZrO$_3$ and ZrO$_2$, respectively. The dashed lines depict the valence and conduction band edges. Here, the VBM and CBM are set to the bulk values of CsPbI$_3$ calculated with a hybrid functional including SOC.} 
\label{alignment}
\end{figure}

Figure~\ref{alignment} summarizes the level alignments for all interfaces investigated in this work.  Our band structures and LDOS plots (Fig.~S4, S5 and S6 of the SI) show no mid gap states. Table~\ref{offsets} summarizes the band offsets for all interfaces. Upon the inclusion of SOC to the bulk band structure calculation, the CBM is pulled into the gap by $\sim 0.8$ eV while VBM shifts up by $\sim 0.1$ eV, reducing the bulk band gap energy to 1.34 eV. To account for the effect of SOC in our coating-perovskite interface level alignments, we shifted the VBM and CBM of the bulk LDOS by $0.1$ and $-0.8$ eV respectively. The band structures and LDOS of all interfaces in this work are shown in Fig.~S4 and S5 of SI respectively.

\begin{table}[!htp]
\caption{Valence (VB) and conduction band (CB) offsets (in eV) at the coating-perovskite interfaces.} \label{offsets}\vspace{-1.5em}
\begin{tabular}{c@{\hspace{1em}}c@{\hspace{1em}}c@{\hspace{2em}}c@{\hspace{1em}} c@{\hspace{2em}}c@{\hspace{1em}}c@{\hspace{2em}}c@{\hspace{1em}} c@{\hspace{1em}}r} 
\\ 
& \multicolumn{8}{c}{Band offsets [eV]} 
\\\hline
& \multicolumn{2}{c}{clean} & \multicolumn{2}{c} {~$i_{\text{PbI}_2^{}}$} & \multicolumn{2}{c} {~$i_{2\text{PbI}_2^{}}$} & \multicolumn{2}{c} {~$i_{4\text{CsI}}$}
\\\hline
 & ~ VB & ~ CB & ~ VB & ~ CB & ~ VB & ~ CB & ~ VB & ~ CB \\\hline\hline
 SrZrO$_3$ & ~1.46& ~3.05 & ~2.62 & ~2.33& ~2.61 & ~1.96& ~2.05& ~2.65\\
 
 ZrO$_2$ & ~2.56 & ~2.12& ~3.71& ~1.12& ~3.72 & ~1.11 & ~3.33 & ~1.35\\
 
 ZnO & ~1.69 & ~2.03& ~1.96 & ~2.48 &~2.16 & ~2.15 & ~1.43 & ~2.28 \\
 \hline\hline
\end{tabular}
\end{table}

In all cases, we observe a type I level alignment (straddling-gap). Of all three coatings, ZrO$_2$-based interfaces show the largest VB offsets with ZrO$_2$@i$_{\text{PbI}_2^{}}$ exhibiting the largest. Concomitantly, the CB offsets are the smallest. Conversely, ZnO-based interfaces (except for ZnO@i$_{\text{PbI}_2^{}}$), show larger CB offsets. Interestingly, we observe a mixed trend in SrZrO$_3$-based interfaces. Specifically, SrZrO$_3$@clean and SrZrO$_3$@i$_{4\text{CsI}}$ show larger CB to VB offsets while the opposite is seen in SrZrO$_3$@i$_{\text{PbI}_2^{}}$ and SrZrO$_3$@i$_{2\text{PbI}_2^{}}$.

\section{Discussion}\label{discussion}

Despite noticeable atomic displacement in the topmost layers of the substrates (Fig.~\ref{structure}), Fig.~S4 and S5 of the SI show no electronic states in the band gap. The absence of such gap states is beneficial for devices since they could act as nonradiative recombination sites. We cannot, however, exclude the presence of structural defects in real devices that introduce such gap states. 

The different band  alignments in Fig.~\ref{alignment} show that the interface can be engineered to enhance charge collection or blocking. Studies have shown that ETLs (and HTLs) with wide band gap, small CB (VB), and large VB (CB) offsets to the substrates have the potential to efficiently fulfill exciton confinement and hole-blocking (electron-blocking) functions in PSCs \cite{Lee2012, Lian2018}. ZnO is known to be a wide band gap ($>$3 eV) $n$-type semiconductor that has been widely explored in optoelectronics \cite{Sun2011, Ro2016, Liang2015, Yu2015, Tong2016, Zheng2019,Liang2015}. SrZrO$_3$ is also an intrinsic perovkite semiconducting material which means, it has the tendency to transport both electrons and holes. Conversely, ZrO$_2$ is a well known insulating material that has been used as protective coating in optoelectronics \cite{Escobar17, LiY2018}.

Our results indicate that ZrO$_2$ might not only act as insulating layer on CsPbI$_3$, but could be engineered to be an ETL. The CB offsets are the lowest we observe for all interfaces in this work and additional iodine in the form of PbI$_2$ or CsI reduces the CB offsets considerably compared to the clean interface. This suggests that further interface modifications might lower the CB offsets sufficiently for ZrO$_2$ to become an ETL.

 \section{Conclusion}\label{conclusion}
In summary, we have successfully studied the interactions of the coating materials ZnO, SrZrO$3$ and ZrO$_2$ on four reconstructed CsPbI$_3$ surface models (clean, $i_{\text{PbI}_2^{}}$, $i_{2\text{PbI}_2^{}}$ and $i_{4\text{CsI}}$) by combining a machine learning based structural search method and DFT. Our optimized structures show strong bonding between the coatings and the substrates at the interfaces. 
Despite the changes in the atomic positions at the topmost layers of our substrates, our spatially resolved local density of states analysis shows no mid gap states which is good for transport properties across the interfaces. We further observed that both the VB and CB offsets for all coatings are large. ZrO$_2$ exhibits the smallest CB offset and could potentially, with the right interface engineering, serve as an ETL. Our current and previous studies serve as a starting point for future work on surface adsorbates, defects and interface engineering of PSCs.  

\section{Computational details}\label{method}

Data acquisitions that serve as input to BOSS are the binding energies ($E_\text{b}$) of single-point calculations at the relative perovskite-coating shift ($x$,$y$) suggested by the BOSS acquisition function. The atomic structures of coating and perovskite are kept fixed. With each additional sampled configuration, Gaussian process models for the energetics are fitted with an uninformative prior on the mean and the model hyperparameters are optimized following the standard procedure of maximising the log marginal likelihood. We employed the exploration-biased eLCB acquisition function. The procedure is iterated until the BOSS surrogate model converges (20 iterations of the coatings on clean and $i_{4\text{CsI}}$, 50 for coatings on surfaces with added PbI$_2$ units). We monitor convergence by tracking the minima of E$_b$ within [$-3$:0] eV. We then extract the global minimum from the BOSS surrogate model and use the structural model at the corresponding ($x$,$y$) coordinates as input for DFT geometry optimization. 
 
 \subsection{Interface structural information} 
We used BOSS and DFT to search for the optimal configurations of our perovskite-coating interfaces. For all our coating materials, we used a $3\times3\times3$ supercell, which is close to being commensurate with a $2\times2$ CsPbI$_3$ substrate supercell. To facilitate DFT interface calculations with periodic boundary conditions, we adjusted the lattice constants  of the coatings (ZnO; $a=b=c=13.02~\mathrm{\AA}$, SrZrO$_3$; $a=b=c=12.59~\mathrm{\AA}$ and ZrO$_2$; $a=b=12.19~\mathrm{\AA}$, $c= 21.57~\mathrm{\AA}$) slightly such that 3 coating unit cells fit exactly onto 2 CsPbI$_3$ unit cells (i.e., are 12.46 {\AA} long).  In this work, we chose the cubic structure of ZnO due to its commensurabililty with the substrate even though the wurtzite structure is the most common polymorph used in optoelectronics. However, studies have shown \cite{Narotsky2004, Demiroglu2013, Wu2011, Freeman2006,Bieniek2015} that in real device engineering, the reduction in size of inorganic materials to the nanoscale induces different structural ordering relative to the most stable bulk polymorph. The perovskite surface models used as substrate in this work are the most stable reconstructed surfaces (clean, $i_{\text{PbI}_2^{}}$,  $i_{2\text{PbI}_2^{}}$ and  $i_{4\text{CsI}}$) of $\upalpha$-${\text{CsPbI}_2^{}}$ from our previous work \cite{Seidu2021}.

\subsection{Boundary conditions for structural search}
\begin{figure}[!ht]
\centering
\includegraphics[width=\linewidth]{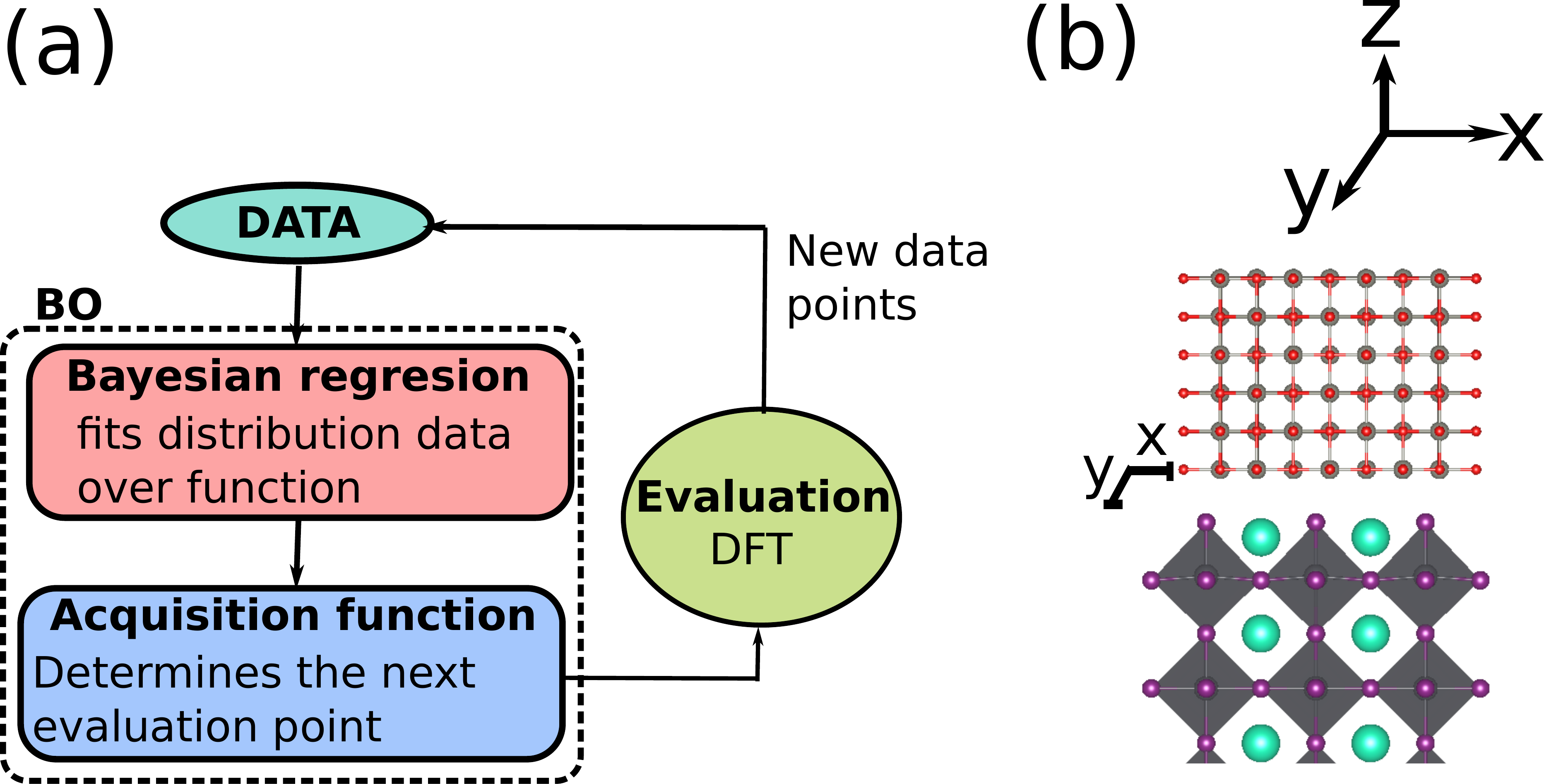}\vspace{0.1em}
\caption{Workflow of the BOSS structural search and an example of its performance. (a) Principles governing the build up of surrogate model by BOSS. (b) Registry of the structure under investigation from which the binding energy is determined by varying $x$ and $y$.} 
\label{workflow}
\end{figure}

Figure~\ref{workflow}a shows the three-step approach for the structural search with BOSS and DFT. Figure~\ref{workflow}b depicts the registry for ZnO on the CsI-T surface model. $x$ and $y$ are the translations of the coating in the $x$-$y$ plane away from an arbitrary initial origin ([0,0]) which corresponds to a search domain $x=y\in$[0.0-2.08] {\AA} for coatings on the clean and $i_{4\text{CsI}}$ substrates. For coatings on substrates with added PbI$_2$ units, the search domain correspond to $x=y\in$[0.0-4.15] {\AA}. By symmetry and with our choice of origin, the translations of the coating in $x$ and $y$ are identical . 
 
In all our structural models, we kept the distance between the coating and the perovskite along the $z$-axis constant at $\sim 4$ {\AA} based on preliminary investigations with a 3-dimensional search space. In our case, the z-direction is not so important since our tests show it does not affect the shape or features of the 2D energy landscape and the final relaxation corrects the bonding of the atoms at the interface.
 
\subsection{DFT and band alignment computations}
 To overcome numerical error from BOSS, which does not enforce symmetry,  we averaged the optimal $x$ and $y$ translations at the local minima  obtained from the BOSS/DFT run before relaxation. 
By fixing the bulk units (all layers below the topmost CsI and PbI$_2$ units) of the substrates, we then relaxed these structures, including the out-of-plane distance $z$, and calculated the spatially resolved local density of states (LDOS) using the hybrid Heyd-Scuseria-Ernzerhof (HSE06, simplified as HSE in this paper) \cite{Heyd2003} functional. From the HSE calculations, we deduced the band alignments at the coating-perovskite interfaces. 

We used the PBEsol \cite{perdew08} functional  with tier-1 basis sets for the single point BOSS/DFT calculations and tier-2 basis sets for the structural relaxation (as implemented in the \textsc{fhi-aims} code \cite{Levchenko/etal:2015, Heyd2003}). In all cases, we used a Gamma-centered $4\times4\times1$ $k$-point grid. We also included  dipole corrections \cite{Neugebauer92} and a vacuum size of $\sim40~\mathrm{\AA}$ to avoid dipole interactions between neighboring slabs. 

In our HSE calculations, we used the standard range-separation parameter $\omega$=0.11 $\text{bohr}^{-1}$ but adjusted the amount of exact exchange ($\alpha$) to 0.55. We also included spin-orbit coupling to our bulk band structure calculations to establish the effect of band splitting on the electronic properties. The $\alpha$ value was obtained by fitting our HSE+SOC band gap to the $GW$ band gap (E$_{\text{g}}$= 1.48 eV) of the cubic ($\upalpha$) CsPbI$_3$ structure reported in Ref.~\onlinecite{Marronnier2018} (see SI for details).

\section{Author information}
\textbf{Corresponding Authors}\\
\textbf{Azimatu Fangnon --}~{Department of Applied Physics, Aalto University, FI-00076 AALTO,Finland}; \\\orcid{0000-0002-9416-7367}\hyperlink{orcid.org/0000-0002-9416-7367}{orcid.org/0000-0002-9416-7367};\\ \hyperlink{Email:~azimatu.fangnon@aalto.fi}{\textcolor{black}{Email:}~azimatu.fangnon@aalto.fi}\\\\
\textbf{Authors}\\
\textbf{Marc Dvorak --}~{Department of Applied Physics, Aalto University, FI-00076 AALTO, Finland};\\ \orcid{0000-0001-9653-2674}\hyperlink{orcid.org/0000-0001-9653-2674}{orcid.org/0000-0001-9653-2674};
\hyperlink{Email:~marc.dvorak@aalto.fi}{\textcolor{black}{Email:}~marc.dvorak@aalto.fi}\\
\textbf{Milica~Todorovic -}~Department of Mechanical and Materials Engineering, University of Turku, FI-02458963 Turku, Finland;\\ 
\orcid{orcid.org/0000-0003-0028-0105}\hyperlink{orcid.org/0000-0003-0028-0105}{orcid.org/0000-0003-0028-0105};\\
\hyperlink{Email:~milica.todorovic@utu.fi}{\textcolor{black}{Email:}~milica.todorovic@utu.fi}\\
\textbf{Ville Havu --}~{Department of Applied Physics, Aalto University, FI-00076 AALTO, Finland};\\ 
\orcid{}\hyperlink{}{};\\
\hyperlink{Email:~ville.havu@aalto.fi}{\textcolor{black}{Email:}~ville.havu@aalto.fi}\\
\textbf{Jingrui~Li -}~Electronic Materials Research Laboratory, Key Laboratory of the Ministry of Education \& International Center for Dielectric Research, School of Electronic Science and Engineering, Xi'an Jiaotong University, Xi'an 710049, China;\\
\orcid{orcid.org/0000-0003-0348-068X}\hyperlink{orcid.org/0000-0003-0348-068X}{orcid.org/0000-0003-0348-068X};\\
\hyperlink{Email:~jingrui.li@xjtu.edu.cn}{\textcolor{black}{Email:}~jingrui.li@xjtu.edu.cn}\\
\textbf{Patrick Rinke --}~Department of Applied Physics, Aalto University, FI-00076 AALTO, Finland;\\
\orcid{0000-0003-1898-723X}\hyperlink{orcid.org/0000-0003-1898-723X}{orcid.org/0000-0003-1898-723X};\\
\hyperlink{Email:~patrick.rinke@aalto.fi}{\textcolor{black}{Email:}~patrick.rinke@aalto.fi}

\begin{acknowledgement}

We acknowledge the computing resources from the CSC-IT Center for Science, the Aalto Science-IT project, and Xi’an Jiaotong University’s HPC Platform. We further acknowledge funding  from the  V\"ais\"al\"a Foundation and the Academy of Finland through its Key Project Funding scheme (305632) and postdoctoral grant no.~316347.

\end{acknowledgement}

\section{Conflict of interest}
The authors have no conflicts to disclose.

\section{Data sharing policy} 
The data that supports the findings of this study will be openly available in Novel Materials Discovery (NOMAD) repository at \cite{Note-NOMAD}.
\begin{suppinfo}
See Supplementary Information for energy landscapes of SrZrO$_3$ and ZrO$_2$  in Fig.~S1. The binding energies of the interfaces before relaxation are also shown in Table.~S1. We present the optimized interface structures of SrZrO$_3$ and ZrO$_2$ in Fig.~S2. Also contained in the SI is a plot of the optimized binding energies of all interfaces as a function of lattice strain (Fig.~S3). The band structures and spatially resolved density of states (LDOS) of all interfaces are presented in Fig.~S4 and S5 respectively. 
\end{suppinfo}

\nocite{References}
\bibliography{achemso-demo}

\providecommand{\latin}[1]{#1}
\makeatletter
\providecommand{\doi}
  {\begingroup\let\do\@makeother\dospecials
  \catcode`\{=1 \catcode`\}=2 \doi@aux}
\providecommand{\doi@aux}[1]{\endgroup\texttt{#1}}
\makeatother
\providecommand*\mcitethebibliography{\thebibliography}
\csname @ifundefined\endcsname{endmcitethebibliography}
  {\let\endmcitethebibliography\endthebibliography}{}
\begin{mcitethebibliography}{76}
\providecommand*\natexlab[1]{#1}
\providecommand*\mciteSetBstSublistMode[1]{}
\providecommand*\mciteSetBstMaxWidthForm[2]{}
\providecommand*\mciteBstWouldAddEndPuncttrue
  {\def\EndOfBibitem{\unskip.}}
\providecommand*\mciteBstWouldAddEndPunctfalse
  {\let\EndOfBibitem\relax}
\providecommand*\mciteSetBstMidEndSepPunct[3]{}
\providecommand*\mciteSetBstSublistLabelBeginEnd[3]{}
\providecommand*\EndOfBibitem{}
\mciteSetBstSublistMode{f}
\mciteSetBstMaxWidthForm{subitem}{(\alph{mcitesubitemcount})}
\mciteSetBstSublistLabelBeginEnd
  {\mcitemaxwidthsubitemform\space}
  {\relax}
  {\relax}

\bibitem[Kim \latin{et~al.}(2012)Kim, Lee, Im, Lee, Moehl, Marchioro, Moon,
  Humphry-Baker, Yum, Moser, Gr{\"a}tzel, and Park]{Kim2012}
Kim,~H.-S.; Lee,~C.-R.; Im,~J.-H.; Lee,~K.-B.; Moehl,~T.; Marchioro,~A.;
  Moon,~S.-J.; Humphry-Baker,~R.; Yum,~J.-H.; Moser,~J.~E.; Gr{\"a}tzel,~M.;
  Park,~N.-G. {Lead Iodide Perovskite Sensitized All-Solid-State Submicron Thin
  Film Mesoscopic Solar Cell with Efficiency Exceeding} 9{\%}. \emph{Sci. Rep.}
  \textbf{2012}, \emph{2}, 1\relax
\mciteBstWouldAddEndPuncttrue
\mciteSetBstMidEndSepPunct{\mcitedefaultmidpunct}
{\mcitedefaultendpunct}{\mcitedefaultseppunct}\relax
\EndOfBibitem
\bibitem[Lee \latin{et~al.}(2012)Lee, Teuscher, Miyasaka, Murakami, and
  Snaith]{Lee2012}
Lee,~M.~M.; Teuscher,~J.; Miyasaka,~T.; Murakami,~T.~N.; Snaith,~H. {Efficient
  Hybrid Solar Cells Based on Meso-Superstructured Oraganometal Halide
  Perovskites}. \emph{Science} \textbf{2012}, \emph{338}, 643\relax
\mciteBstWouldAddEndPuncttrue
\mciteSetBstMidEndSepPunct{\mcitedefaultmidpunct}
{\mcitedefaultendpunct}{\mcitedefaultseppunct}\relax
\EndOfBibitem
\bibitem[Min \latin{et~al.}(2021)Min, Lee, Kim, Kim, Lee, Kim, Paik, Kim, Kim,
  Kim, Shin, and Seok]{MinH2021}
Min,~H.; Lee,~D.~Y.; Kim,~J.; Kim,~G.; Lee,~K.~S.; Kim,~J.; Paik,~M.~J.;
  Kim,~Y.~K.; Kim,~K.~S.; Kim,~M.~G.; Shin,~T.~J.; Seok,~S.~I. {Perovskite
  Solar Cells with Atomically Coherent Interlayers on {SnO$_2^{}$} Electrodes}.
  \emph{Nature} \textbf{2021}, \emph{598}, 444\relax
\mciteBstWouldAddEndPuncttrue
\mciteSetBstMidEndSepPunct{\mcitedefaultmidpunct}
{\mcitedefaultendpunct}{\mcitedefaultseppunct}\relax
\EndOfBibitem
\bibitem[Chen \latin{et~al.}(2018)Chen, Bai, Lyu, Yun, Hao, and
  Wang]{ChenP2018}
Chen,~P.; Bai,~Y.; Lyu,~M.; Yun,~J.-H.; Hao,~M.; Wang,~L. {Progress and
  Perspective in Low-Dimensional Metal Halide Perovskites for Optoelectronic
  Applications}. \emph{Sol. RRL} \textbf{2018}, \emph{2}, 1700186\relax
\mciteBstWouldAddEndPuncttrue
\mciteSetBstMidEndSepPunct{\mcitedefaultmidpunct}
{\mcitedefaultendpunct}{\mcitedefaultseppunct}\relax
\EndOfBibitem
\bibitem[Zhang \latin{et~al.}(2020)Zhang, Kuang, and Wu]{ZhangC2020}
Zhang,~C.; Kuang,~D.-B.; Wu,~W.-Q. {A Review of Diverse Halide Perovskite
  Morphologies for Efficient Optoelectronic Applications}. \emph{Small Methods}
  \textbf{2020}, \emph{4}, 1900662\relax
\mciteBstWouldAddEndPuncttrue
\mciteSetBstMidEndSepPunct{\mcitedefaultmidpunct}
{\mcitedefaultendpunct}{\mcitedefaultseppunct}\relax
\EndOfBibitem
\bibitem[Dong \latin{et~al.}(2020)Dong, Zhang, Liu, Yao, and Zhao]{DongH2020}
Dong,~H.; Zhang,~C.; Liu,~X.; Yao,~J.; Zhao,~Y.~S. {Materials Chemistry and
  Engineering in Metal Halide Perovskite Lasers}. \emph{Chem. Soc. Rev.}
  \textbf{2020}, \emph{49}, 951\relax
\mciteBstWouldAddEndPuncttrue
\mciteSetBstMidEndSepPunct{\mcitedefaultmidpunct}
{\mcitedefaultendpunct}{\mcitedefaultseppunct}\relax
\EndOfBibitem
\bibitem[Wang \latin{et~al.}(2021)Wang, Liu, Cao, and Wang]{WangY2021}
Wang,~Y.; Liu,~Y.; Cao,~S.; Wang,~J. {A Review on Solution-Processed
  Perovskite/Organic Hybrid Photodetectors}. \emph{J. Mater. Chem. C}
  \textbf{2021}, \emph{9}, 5302\relax
\mciteBstWouldAddEndPuncttrue
\mciteSetBstMidEndSepPunct{\mcitedefaultmidpunct}
{\mcitedefaultendpunct}{\mcitedefaultseppunct}\relax
\EndOfBibitem
\bibitem[Correa-Baena \latin{et~al.}(2017)Correa-Baena, Saliba, Buonassisi,
  Gr\"atzel, Abate, Tress, and Hagfeldt]{CorreaBaena2017}
Correa-Baena,~J.-P.; Saliba,~M.; Buonassisi,~T.; Gr\"atzel,~M.; Abate,~A.;
  Tress,~W.; Hagfeldt,~A. {Promises and Challenges of Perovskite Solar Cells}.
  \emph{Science} \textbf{2017}, \emph{358}, 739\relax
\mciteBstWouldAddEndPuncttrue
\mciteSetBstMidEndSepPunct{\mcitedefaultmidpunct}
{\mcitedefaultendpunct}{\mcitedefaultseppunct}\relax
\EndOfBibitem
\bibitem[Rajagopal \latin{et~al.}(2018)Rajagopal, Yao, and Jen]{Rajagopal2018}
Rajagopal,~A.; Yao,~K.; Jen,~A.~K.-Y. {Toward Perovskite Solar Cell
  Commercialization: A Perspective and Research Roadmap Based on Interfacial
  Engineering}. \emph{Adv. Mater.} \textbf{2018}, \emph{30}, 1800455\relax
\mciteBstWouldAddEndPuncttrue
\mciteSetBstMidEndSepPunct{\mcitedefaultmidpunct}
{\mcitedefaultendpunct}{\mcitedefaultseppunct}\relax
\EndOfBibitem
\bibitem[Li \latin{et~al.}(2020)Li, Wei, and Ning]{LiH2020}
Li,~H.; Wei,~Q.; Ning,~Z. {Toward High Efficiency Tin Perovskite Solar Cells: A
  Perspective}. \emph{Appl. Phys. Lett.} \textbf{2020}, \emph{117},
  060502\relax
\mciteBstWouldAddEndPuncttrue
\mciteSetBstMidEndSepPunct{\mcitedefaultmidpunct}
{\mcitedefaultendpunct}{\mcitedefaultseppunct}\relax
\EndOfBibitem
\bibitem[Niu \latin{et~al.}(2014)Niu, Li, Meng, Wang, Dong, and Qiu]{NiuG14}
Niu,~G.; Li,~W.; Meng,~F.; Wang,~L.; Dong,~H.; Qiu,~Y. {{Study on the Stability
  of CH$_3$NH$_3$PbI$_3$ Films and the Effect of Post-modification by
  Aluminumoxide in All-solid-state Hybrid Solar Cells}}. \emph{J. Mater. Chem.
  A} \textbf{2014}, \emph{2}, 705--710\relax
\mciteBstWouldAddEndPuncttrue
\mciteSetBstMidEndSepPunct{\mcitedefaultmidpunct}
{\mcitedefaultendpunct}{\mcitedefaultseppunct}\relax
\EndOfBibitem
\bibitem[Niu \latin{et~al.}(2015)Niu, Guo, and Wang]{NiuG15}
Niu,~G.; Guo,~X.; Wang,~L. {Review of Recent Progress in Chemical Stability of
  Perovskite Solar Cells}. \emph{J. Mater. Chem. A} \textbf{2015}, \emph{3},
  8970\relax
\mciteBstWouldAddEndPuncttrue
\mciteSetBstMidEndSepPunct{\mcitedefaultmidpunct}
{\mcitedefaultendpunct}{\mcitedefaultseppunct}\relax
\EndOfBibitem
\bibitem[Huang \latin{et~al.}(2017)Huang, Tan, Lund, and Zhou]{HuangJ17}
Huang,~J.; Tan,~S.; Lund,~P.~D.; Zhou,~H. {{Impact of H$_2$O on
  Organic–inorganic Hybrid Perovskite Solar Cells}}. \emph{Energy Environ.
  Sci.} \textbf{2017}, \emph{10}, 2284\relax
\mciteBstWouldAddEndPuncttrue
\mciteSetBstMidEndSepPunct{\mcitedefaultmidpunct}
{\mcitedefaultendpunct}{\mcitedefaultseppunct}\relax
\EndOfBibitem
\bibitem[Kim \latin{et~al.}(2017)Kim, Jang, Yoon, Jeong, Park, Walker, Jeon,
  Jo, Yoon, Kim, Baek, K, and Kim]{KimGH17}
Kim,~G.-H.; Jang,~H.; Yoon,~Y.~J.; Jeong,~J.; Park,~S.~Y.; Walker,~B.;
  Jeon,~I.-Y.; Jo,~Y.; Yoon,~H.; Kim,~M.; Baek,~J.-B.; K,~D.~S.; Kim,~J.~Y.
  {Fluorine Functionalized Graphene Nano Platelets for Highly Stable Inverted
  Perovskite Solar Cells}. \emph{Nano Lett.} \textbf{2017}, \emph{17},
  6385\relax
\mciteBstWouldAddEndPuncttrue
\mciteSetBstMidEndSepPunct{\mcitedefaultmidpunct}
{\mcitedefaultendpunct}{\mcitedefaultseppunct}\relax
\EndOfBibitem
\bibitem[Mesquita \latin{et~al.}(2018)Mesquita, Andrade, and
  Mendes]{Mesquita18}
Mesquita,~I.; Andrade,~L.; Mendes,~A. {Perovskite Solar Cells: Materials,
  Configurations and Stability}. \emph{Renew. Sust. Energy Rev.} \textbf{2018},
  \emph{82}, 2471\relax
\mciteBstWouldAddEndPuncttrue
\mciteSetBstMidEndSepPunct{\mcitedefaultmidpunct}
{\mcitedefaultendpunct}{\mcitedefaultseppunct}\relax
\EndOfBibitem
\bibitem[Li \latin{et~al.}(2018)Li, Yuan, Ling, Zhang, Yang, Cheung, Ho, Gao,
  and Ma]{LiF18}
Li,~F.; Yuan,~J.; Ling,~X.; Zhang,~Y.; Yang,~Y.; Cheung,~S.~H.; Ho,~C. H.~Y.;
  Gao,~X.; Ma,~W. {A Universal Strategy to Utilize Polymeric Semiconductors for
  Perovskite Solar Cells with Enhanced Efficiency and Longevity}. \emph{Adv.
  Funct. Mater.} \textbf{2018}, \emph{18}, 1706377\relax
\mciteBstWouldAddEndPuncttrue
\mciteSetBstMidEndSepPunct{\mcitedefaultmidpunct}
{\mcitedefaultendpunct}{\mcitedefaultseppunct}\relax
\EndOfBibitem
\bibitem[Shao \latin{et~al.}(2015)Shao, Yuan, and Huang]{Shao2015}
Shao,~Y.; Yuan,~Y.; Huang,~J. {Correlation of Energy Disorder and Open-circuit
  Voltage in Hybrid Perovskite Solar Cells}. \emph{Nat. Energy} \textbf{2015},
  \emph{1}, 15001\relax
\mciteBstWouldAddEndPuncttrue
\mciteSetBstMidEndSepPunct{\mcitedefaultmidpunct}
{\mcitedefaultendpunct}{\mcitedefaultseppunct}\relax
\EndOfBibitem
\bibitem[Leijtens \latin{et~al.}(2013)Leijtens, Lim, Teuscher, Park, and
  Snaith]{Leijtens2013}
Leijtens,~T.; Lim,~J.; Teuscher,~J.; Park,~T.; Snaith,~H.~J. {Charge Density
  Dependent Mobility of Organic Hole-Transporters and Mesoporous TiO$_2$
  Determined by Transient Mobility Spectroscopy: Implications to Dye-Sensitized
  and Organic Solar Cells}. \emph{Adv. Mater.} \textbf{2013}, \emph{25},
  3227--3233\relax
\mciteBstWouldAddEndPuncttrue
\mciteSetBstMidEndSepPunct{\mcitedefaultmidpunct}
{\mcitedefaultendpunct}{\mcitedefaultseppunct}\relax
\EndOfBibitem
\bibitem[Im \latin{et~al.}(2011)Im, Lee, Lee, Park, and Park]{Im2011}
Im,~J.-H.; Lee,~C.-R.; Lee,~J.-W.; Park,~S.-W.; Park,~N.-G. {6.5\% Efficient
  Perovskite Quantum-Dot-Sensitized Solar Cell}. \emph{Nanoscale}
  \textbf{2011}, \emph{3}, 4088\relax
\mciteBstWouldAddEndPuncttrue
\mciteSetBstMidEndSepPunct{\mcitedefaultmidpunct}
{\mcitedefaultendpunct}{\mcitedefaultseppunct}\relax
\EndOfBibitem
\bibitem[Burschka \latin{et~al.}(2013)Burschka, Pellet, Moon, Humphry-Baker,
  Gao, Nazeeruddin, and Gr{\"a}tzel]{Burschka2013}
Burschka,~J.; Pellet,~N.; Moon,~S.-J.; Humphry-Baker,~R.; Gao,~P.;
  Nazeeruddin,~M.~K.; Gr{\"a}tzel,~M. {Sequential Deposition as a Route to
  High-Performance Perovskite-Sensitized Solar Cells}. \emph{Nature}
  \textbf{2013}, \emph{499}, 316\relax
\mciteBstWouldAddEndPuncttrue
\mciteSetBstMidEndSepPunct{\mcitedefaultmidpunct}
{\mcitedefaultendpunct}{\mcitedefaultseppunct}\relax
\EndOfBibitem
\bibitem[Burschka \latin{et~al.}(2011)Burschka, Dualeh, Florian~Kessler,
  Cevey-Ha, Yi, Nazeeruddin, and Gr{\"a}tzel]{Burschka2011}
Burschka,~J.; Dualeh,~A.; Florian~Kessler,~E.~B.; Cevey-Ha,~N.-L.; Yi,~C.;
  Nazeeruddin,~M.~K.; Gr{\"a}tzel,~M.
  {Tris(2-(1H-pyrazol-1-yl)pyridine)cobalt(III) as p-Type Dopant for Organic
  Semiconductors and Its Application in Highly Efficient Solid-State
  Dye-Sensitized Solar Cells}. \emph{J. Am. Chem. Soc.} \textbf{2011},
  \emph{133}, 18042--18045\relax
\mciteBstWouldAddEndPuncttrue
\mciteSetBstMidEndSepPunct{\mcitedefaultmidpunct}
{\mcitedefaultendpunct}{\mcitedefaultseppunct}\relax
\EndOfBibitem
\bibitem[Saliba \latin{et~al.}(2016)Saliba, Orlandi, Matsui, Aghazada,
  Cavazzini, Correa-Baena, Gao, Scopelliti, Mosconi, Dahmen, Angelis, Abate,
  Hagfeldt, Pozzi, Gr\"atzel, and Nazeeruddin]{Saliba16b}
Saliba,~M.; Orlandi,~S.; Matsui,~T.; Aghazada,~S.; Cavazzini,~M.;
  Correa-Baena,~J.-P.; Gao,~P.; Scopelliti,~R.; Mosconi,~E.; Dahmen,~K.-H.;
  Angelis,~F.~D.; Abate,~A.; Hagfeldt,~A.; Pozzi,~G.; Gr\"atzel,~M.;
  Nazeeruddin,~M.~K. {A Molecularly Engineered Hole-transporting Material for
  Efficient Perovskite Solar Cells}. \emph{Nat. Energy} \textbf{2016},
  \emph{1}, 15017\relax
\mciteBstWouldAddEndPuncttrue
\mciteSetBstMidEndSepPunct{\mcitedefaultmidpunct}
{\mcitedefaultendpunct}{\mcitedefaultseppunct}\relax
\EndOfBibitem
\bibitem[Schmidt \latin{et~al.}(2014)Schmidt, Perteg{\'a},
  Gonz{\'a}lez-Carrero, Malinkiewicz, Agouram, Espallargas, Bolink, Galian, and
  P{\'e}rez-Prieto]{SchmidtL14}
Schmidt,~L.~C.; Perteg{\'a},~A.; Gonz{\'a}lez-Carrero,~S.; Malinkiewicz,~O.;
  Agouram,~S.; Espallargas,~G.~M.; Bolink,~H.~J.; Galian,~R.~E.;
  P{\'e}rez-Prieto,~J. {Nontemplate Synthesis of
  $\text{CH}_{3}^{}\text{NH}_{3}^{}\text{PbBr}_{3}^{}$ Perovskite
  Nanoparticles}. \emph{J. Am. Chem. Soc.} \textbf{2014}, \emph{136}, 850\relax
\mciteBstWouldAddEndPuncttrue
\mciteSetBstMidEndSepPunct{\mcitedefaultmidpunct}
{\mcitedefaultendpunct}{\mcitedefaultseppunct}\relax
\EndOfBibitem
\bibitem[Gonz{\'a}lez-Carrero \latin{et~al.}(2015)Gonz{\'a}lez-Carrero, Galian,
  and P{\'e}rez-Prieto]{SoranyelG15}
Gonz{\'a}lez-Carrero,~S.; Galian,~R.~E.; P{\'e}rez-Prieto,~J. {Maximizing the
  Emissive Properties of $\text{CH}_{3}^{}\text{NH}_{3}^{}\text{PbBr}_{3}^{}$
  Perovskite Nanoparticles}. \emph{J. Mater. Chem. A} \textbf{2015}, \emph{3},
  9187--9193\relax
\mciteBstWouldAddEndPuncttrue
\mciteSetBstMidEndSepPunct{\mcitedefaultmidpunct}
{\mcitedefaultendpunct}{\mcitedefaultseppunct}\relax
\EndOfBibitem
\bibitem[Dong \latin{et~al.}(2019)Dong, Xi, Zuo, Li, Yang, Wang, Yu, Ma, Ran,
  Gao, Jiao, Xu, Lei, Wei, Yuan, Zhang, Shi, Hou, and Wu]{Dong19}
Dong,~H.; Xi,~J.; Zuo,~L.; Li,~J.; Yang,~Y.; Wang,~D.; Yu,~Y.; Ma,~L.; Ran,~C.;
  Gao,~W.; Jiao,~B.; Xu,~J.; Lei,~T.; Wei,~F.; Yuan,~F.; Zhang,~L.; Shi,~Y.;
  Hou,~X.; Wu,~Z. {Conjugated Molecules ``Bridge``: Functional Ligand Toward
  Highly Efficient and Long-term Stable Perovskite Solar Cell}. \emph{Adv.
  Funct. Mater.} \textbf{2019}, \emph{29}, 1808119\relax
\mciteBstWouldAddEndPuncttrue
\mciteSetBstMidEndSepPunct{\mcitedefaultmidpunct}
{\mcitedefaultendpunct}{\mcitedefaultseppunct}\relax
\EndOfBibitem
\bibitem[Chen \latin{et~al.}(2020)Chen, Li, Li, and Li]{ChenW2020}
Chen,~W.; Li,~X.; Li,~Y.; Li,~Y. A Review: Crystal Growth for High-performance
  All-inorganic Perovskite Solar Cells. \emph{Energy Environ. Sci.}
  \textbf{2020}, \emph{13}, 1971\relax
\mciteBstWouldAddEndPuncttrue
\mciteSetBstMidEndSepPunct{\mcitedefaultmidpunct}
{\mcitedefaultendpunct}{\mcitedefaultseppunct}\relax
\EndOfBibitem
\bibitem[Liu \latin{et~al.}(2020)Liu, Yang, Syzgantseva, Ding, Syzgantseva,
  Zhang, Asiri, Dai, and Nazeeruddin]{LiuC2020}
Liu,~C.; Yang,~Y.; Syzgantseva,~O.~A.; Ding,~Y.; Syzgantseva,~M.~A.; Zhang,~X.;
  Asiri,~A.~M.; Dai,~S.; Nazeeruddin,~M.~K. {$\upalpha$-CsPbI$_3^{}$ Bilayers
  via One-Step Deposition for Efficient and Stable All-Inorganic Perovskite
  Solar Cells}. \emph{Adv. Mater.} \textbf{2020}, \emph{32}, 2002632\relax
\mciteBstWouldAddEndPuncttrue
\mciteSetBstMidEndSepPunct{\mcitedefaultmidpunct}
{\mcitedefaultendpunct}{\mcitedefaultseppunct}\relax
\EndOfBibitem
\bibitem[Zhang \latin{et~al.}(2020)Zhang, Yuan, Xi, Jiao, Dong, Li, and
  Wu]{ZhangL2020}
Zhang,~L.; Yuan,~F.; Xi,~J.; Jiao,~B.; Dong,~H.; Li,~J.; Wu,~Z. {Suppressing
  Ion Migration Enables Stable Perovskite Light-Emitting Diodes with
  All-Inorganic Strategy}. \emph{Adv. Funct. Mater.} \textbf{2020}, \emph{30},
  2001834\relax
\mciteBstWouldAddEndPuncttrue
\mciteSetBstMidEndSepPunct{\mcitedefaultmidpunct}
{\mcitedefaultendpunct}{\mcitedefaultseppunct}\relax
\EndOfBibitem
\bibitem[Matteocci \latin{et~al.}(2016)Matteocci, Cin\`a, Lamanna, Cacovich,
  Divitini, Midgley, Ducati, and di~Carlo]{Matteocci16}
Matteocci,~F.; Cin\`a,~L.; Lamanna,~E.; Cacovich,~S.; Divitini,~G.;
  Midgley,~P.~A.; Ducati,~C.; di~Carlo,~A. {Encapsulation for Long-term
  Stability Enhancement of Perovskite Solar Cells}. \emph{Nano Energy}
  \textbf{2016}, \emph{30}, 162\relax
\mciteBstWouldAddEndPuncttrue
\mciteSetBstMidEndSepPunct{\mcitedefaultmidpunct}
{\mcitedefaultendpunct}{\mcitedefaultseppunct}\relax
\EndOfBibitem
\bibitem[Cheacharoen \latin{et~al.}(2018)Cheacharoen, Rolston, Harwood, Bush,
  Dauskardt, and McGehee]{Cheacharoen18a}
Cheacharoen,~R.; Rolston,~N.; Harwood,~D.; Bush,~K.~A.; Dauskardt,~R.~H.;
  McGehee,~M.~D. {Design and Understanding of Encapsulated Perovskite Solar
  Cells to withstand Temperature Cycling}. \emph{Energy Environ. Sci.}
  \textbf{2018}, \emph{11}, 144\relax
\mciteBstWouldAddEndPuncttrue
\mciteSetBstMidEndSepPunct{\mcitedefaultmidpunct}
{\mcitedefaultendpunct}{\mcitedefaultseppunct}\relax
\EndOfBibitem
\bibitem[Cheacharoen \latin{et~al.}(2018)Cheacharoen, Boyd, Burkhard, Leijtens,
  Raiford, Bush, Bent, and McGehee]{Cheacharoen18b}
Cheacharoen,~R.; Boyd,~C.~C.; Burkhard,~G.~F.; Leijtens,~T.; Raiford,~J.~A.;
  Bush,~K.~A.; Bent,~S.~F.; McGehee,~M.~D. {Encapsulating Perovskite Solar
  Cells to withstand Damp Heat and Thermal Cycling}. \emph{Sustainable Energy
  Fuels} \textbf{2018}, \emph{2}, 2398\relax
\mciteBstWouldAddEndPuncttrue
\mciteSetBstMidEndSepPunct{\mcitedefaultmidpunct}
{\mcitedefaultendpunct}{\mcitedefaultseppunct}\relax
\EndOfBibitem
\bibitem[Seidu \latin{et~al.}(2019)Seidu, Himanen, Li, and Rinke]{Seidu19}
Seidu,~A.; Himanen,~L.; Li,~J.; Rinke,~P. {Database-driven High-throughput
  Study of Coating Materials for Hybrid Perovskites}. \emph{New J. Phys.}
  \textbf{2019}, \emph{21}, 083018\relax
\mciteBstWouldAddEndPuncttrue
\mciteSetBstMidEndSepPunct{\mcitedefaultmidpunct}
{\mcitedefaultendpunct}{\mcitedefaultseppunct}\relax
\EndOfBibitem
\bibitem[Lian \latin{et~al.}(2018)Lian, Lu, Niu, Zeng, and Zha]{Lian2018}
Lian,~J.; Lu,~B.; Niu,~F.; Zeng,~P.; Zha,~X. {Electron-Transport Materials in
  Perovskite Solar Cells}. \emph{Small {Methods}} \textbf{2018}, \emph{2},
  1800082\relax
\mciteBstWouldAddEndPuncttrue
\mciteSetBstMidEndSepPunct{\mcitedefaultmidpunct}
{\mcitedefaultendpunct}{\mcitedefaultseppunct}\relax
\EndOfBibitem
\bibitem[Tong \latin{et~al.}(2016)Tong, Lin, Wu, and Wang]{Tong2016}
Tong,~X.; Lin,~F.; Wu,~J.; Wang,~Z.~M. High Performance Perovskite Solar Cells.
  \emph{Adv. Sci.} \textbf{2016}, \emph{3}, 1500201\relax
\mciteBstWouldAddEndPuncttrue
\mciteSetBstMidEndSepPunct{\mcitedefaultmidpunct}
{\mcitedefaultendpunct}{\mcitedefaultseppunct}\relax
\EndOfBibitem
\bibitem[Wang \latin{et~al.}(2019)Wang, Wan, Ding, Hu, and Wang]{Wang2019}
Wang,~Y.; Wan,~J.; Ding,~J.; Hu,~J.-S.; Wang,~D. {A Rutile TiO$_2$ Electron
  Transport Layer for the Enhancement of Charge Collection for Efficient
  Perovskite Solar Cells}. \emph{Angew. Chem. Int. Ed.} \textbf{2019},
  \emph{58}, 9414--9418\relax
\mciteBstWouldAddEndPuncttrue
\mciteSetBstMidEndSepPunct{\mcitedefaultmidpunct}
{\mcitedefaultendpunct}{\mcitedefaultseppunct}\relax
\EndOfBibitem
\bibitem[Anaraki \latin{et~al.}(2016)Anaraki, Kermanpur, Steier, Domanski,
  Matsui, Tress, Saliba, Abate, Gr{\"a}tzel, Hagfeld, and
  Correa-Baena]{Anaraki2016}
Anaraki,~E.~H.; Kermanpur,~A.; Steier,~L.; Domanski,~K.; Matsui,~T.; Tress,~W.;
  Saliba,~M.; Abate,~A.; Gr{\"a}tzel,~M.; Hagfeld,~A.; Correa-Baena,~J.-P.
  {Low-Temperature Solution-Processed Tin Oxide as an Alternative Electron
  Transporting Layer for Efficient Perovskite Solar Cells}. \emph{Energy
  Environ. Sci.} \textbf{2016}, \emph{9}, 3128\relax
\mciteBstWouldAddEndPuncttrue
\mciteSetBstMidEndSepPunct{\mcitedefaultmidpunct}
{\mcitedefaultendpunct}{\mcitedefaultseppunct}\relax
\EndOfBibitem
\bibitem[Jiang \latin{et~al.}(2018)Jiang, Zhang, and You]{Jiang18}
Jiang,~Q.; Zhang,~X.; You,~J. {SnO$_2$: A Wonderful Electron Transport Layer
  for Perovskite Solar Cells}. \emph{Small} \textbf{2018}, \emph{14},
  1801154\relax
\mciteBstWouldAddEndPuncttrue
\mciteSetBstMidEndSepPunct{\mcitedefaultmidpunct}
{\mcitedefaultendpunct}{\mcitedefaultseppunct}\relax
\EndOfBibitem
\bibitem[Xiong \latin{et~al.}(2018)Xiong, Guo, Wen, Liu, Yang, Qin, and
  Fang]{Xiong2018}
Xiong,~L.; Guo,~Y.; Wen,~J.; Liu,~H.; Yang,~G.; Qin,~P.; Fang,~G. {Review on
  the Application of SnO$_2$ in Perovskite Solar Cells}. \emph{Adv. Funct.
  Mater.} \textbf{2018}, \emph{28}, 1802757\relax
\mciteBstWouldAddEndPuncttrue
\mciteSetBstMidEndSepPunct{\mcitedefaultmidpunct}
{\mcitedefaultendpunct}{\mcitedefaultseppunct}\relax
\EndOfBibitem
\bibitem[Lin \latin{et~al.}(2021)Lin, Jones, Yang, Duffy, Li, Zhao, Chi, Wang,
  and Wilson]{Lin2021}
Lin,~L.; Jones,~T.~W.; Yang,~T. C.-J.; Duffy,~N.~W.; Li,~J.; Zhao,~L.; Chi,~B.;
  Wang,~X.; Wilson,~G.~J. {Inorganic Electron Transport Materials in Perovskite
  Solar Cells}. \emph{Adv. Funct. Mater.} \textbf{2021}, \emph{31},
  2008300\relax
\mciteBstWouldAddEndPuncttrue
\mciteSetBstMidEndSepPunct{\mcitedefaultmidpunct}
{\mcitedefaultendpunct}{\mcitedefaultseppunct}\relax
\EndOfBibitem
\bibitem[Christians \latin{et~al.}(2014)Christians, Fung, and
  Kamat]{Christians2014}
Christians,~J.~A.; Fung,~R. C.~M.; Kamat,~P.~V. {An Inorganic Hole Conductor
  for Organo-Lead Halide Perovskite Solar Cells. Improved Hole Conductivity
  with Copper Iodide}. \emph{J. Am. Chem. Soc.} \textbf{2014}, \emph{136},
  758--764\relax
\mciteBstWouldAddEndPuncttrue
\mciteSetBstMidEndSepPunct{\mcitedefaultmidpunct}
{\mcitedefaultendpunct}{\mcitedefaultseppunct}\relax
\EndOfBibitem
\bibitem[Huangfu \latin{et~al.}(2015)Huangfu, Shen, Gongbo~Zhu, Gun, and
  Wang]{Huangfu2015}
Huangfu,~M.; Shen,~Y.; Gongbo~Zhu,~M.~C.,~Kai~Xu; Gun,~F.; Wang,~L. {Copper
  Iodide as Inorganic Hole Conductor for Perovskite Solar Cells with Different
  Thickness of Mesoporous Layer and Hole Transport Layer}. \emph{Appl. Surf.
  Sci.} \textbf{2015}, \emph{357}, 2234--2240\relax
\mciteBstWouldAddEndPuncttrue
\mciteSetBstMidEndSepPunct{\mcitedefaultmidpunct}
{\mcitedefaultendpunct}{\mcitedefaultseppunct}\relax
\EndOfBibitem
\bibitem[Kim \latin{et~al.}(2015)Kim, Liang, Williams, Cho, Chueh, Glaz,
  Ginger, and Jen]{Kim2015}
Kim,~J.~H.; Liang,~P.-W.; Williams,~S.~T.; Cho,~N.; Chueh,~C.-C.; Glaz,~M.~S.;
  Ginger,~D.~S.; Jen,~A. K.-Y. High-Performance and Environmentally Stable
  Planar Heterojunction Perovskite Solar Cells Based on a Solution-Processed
  Copper-Doped Nickel Oxide Hole-Transporting Layer. \emph{Adv. Mater.}
  \textbf{2015}, \emph{27}, 695--701\relax
\mciteBstWouldAddEndPuncttrue
\mciteSetBstMidEndSepPunct{\mcitedefaultmidpunct}
{\mcitedefaultendpunct}{\mcitedefaultseppunct}\relax
\EndOfBibitem
\bibitem[Islam \latin{et~al.}(2017)Islam, Yanagida, Shirai, Nabetani, and
  Miyano]{Islam2017}
Islam,~M.~B.; Yanagida,~M.; Shirai,~Y.; Nabetani,~Y.; Miyano,~K. {An Inorganic
  Hole Conductor for Organo-Lead Halide Perovskite Solar Cells. Improved Hole
  Conductivity with Copper Iodide}. \emph{ACS {Omega}} \textbf{2017}, \emph{2},
  2291--2299\relax
\mciteBstWouldAddEndPuncttrue
\mciteSetBstMidEndSepPunct{\mcitedefaultmidpunct}
{\mcitedefaultendpunct}{\mcitedefaultseppunct}\relax
\EndOfBibitem
\bibitem[Chen \latin{et~al.}(2015)Chen, Wu, Yue, Liu, Zhang, Yang, Chen, Bi,
  Ashraful, Gr{\"a}tzel, and Han]{Chen2015}
Chen,~W.; Wu,~Y.; Yue,~Y.; Liu,~J.; Zhang,~W.; Yang,~X.; Chen,~H.; Bi,~E.;
  Ashraful,~I.; Gr{\"a}tzel,~M.; Han,~L. {Efficient and Stable Large-Area
  Perovskite Solar Cells with Inorganic Charge Extraction Layers}.
  \emph{Science} \textbf{2015}, \emph{350}, 6263\relax
\mciteBstWouldAddEndPuncttrue
\mciteSetBstMidEndSepPunct{\mcitedefaultmidpunct}
{\mcitedefaultendpunct}{\mcitedefaultseppunct}\relax
\EndOfBibitem
\bibitem[Subbiah \latin{et~al.}(2014)Subbiah, Halder, Ghosh, Mahuli, Hodes, and
  Sarkar]{Subbiah2014}
Subbiah,~A.~S.; Halder,~A.; Ghosh,~S.; Mahuli,~N.; Hodes,~G.; Sarkar,~S.~K.
  {Inorganic Hole Conducting Layers for Perovskite-Based Solar Cells}. \emph{J.
  Phys. Chem. Lett.} \textbf{2014}, \emph{5}, 1748--1753\relax
\mciteBstWouldAddEndPuncttrue
\mciteSetBstMidEndSepPunct{\mcitedefaultmidpunct}
{\mcitedefaultendpunct}{\mcitedefaultseppunct}\relax
\EndOfBibitem
\bibitem[Wu \latin{et~al.}(2014)Wu, Bai, Xiang, Yuan, Yang, Cui, Gao, Liu, Jin,
  and Sun]{Wu2014}
Wu,~Z.; Bai,~S.; Xiang,~J.; Yuan,~Z.; Yang,~Y.; Cui,~W.; Gao,~X.; Liu,~Z.;
  Jin,~Y.; Sun,~B. {Efficient Planar Heterojunction Perovskite Solar Cells
  Employing Graphene Oxide as Hole Conductor}. \emph{Nanoscale} \textbf{2014},
  \emph{6}, 10505\relax
\mciteBstWouldAddEndPuncttrue
\mciteSetBstMidEndSepPunct{\mcitedefaultmidpunct}
{\mcitedefaultendpunct}{\mcitedefaultseppunct}\relax
\EndOfBibitem
\bibitem[Shi \latin{et~al.}(2018)Shi, Wu, Dong, Li, Xi, Divitini, Ran, Yuan,
  Zhang, Jiao, Hou, and Wu]{ShiY2018}
Shi,~Y.; Wu,~W.; Dong,~H.; Li,~G.; Xi,~K.; Divitini,~G.; Ran,~C.; Yuan,~F.;
  Zhang,~M.; Jiao,~B.; Hou,~X.; Wu,~Z. {A Strategy for Architecture Design of
  Crystalline Perovskite Light-Emitting Diodes with High Performance}.
  \emph{Adv. Mater.} \textbf{2018}, \emph{30}, 1800251\relax
\mciteBstWouldAddEndPuncttrue
\mciteSetBstMidEndSepPunct{\mcitedefaultmidpunct}
{\mcitedefaultendpunct}{\mcitedefaultseppunct}\relax
\EndOfBibitem
\bibitem[Xu \latin{et~al.}(2020)Xu, Qian, Zhang, Lv, Jin, Zhang, Zheng, Li,
  Chen, and Huang]{XuL2020}
Xu,~L.; Qian,~M.; Zhang,~C.; Lv,~W.; Jin,~J.; Zhang,~J.; Zheng,~C.; Li,~M.;
  Chen,~R.; Huang,~W. {In Situ Construction of Gradient Heterojunction using
  Organic $\text{VO}_x^{}$ Precursor for Efficient and Stable Inverted
  Perovskite Solar Cells}. \emph{Nano Energy} \textbf{2020}, \emph{67},
  104244\relax
\mciteBstWouldAddEndPuncttrue
\mciteSetBstMidEndSepPunct{\mcitedefaultmidpunct}
{\mcitedefaultendpunct}{\mcitedefaultseppunct}\relax
\EndOfBibitem
\bibitem[Yuan \latin{et~al.}(2020)Yuan, Ran, Zhang, Dong, Jiao, Hou, Li, and
  Wu]{YuanF2020}
Yuan,~F.; Ran,~C.; Zhang,~L.; Dong,~H.; Jiao,~B.; Hou,~X.; Li,~J.; Wu,~Z. {A
  Cocktail of Multiple Cations in Inorganic Halide Perovskite Toward Efficient
  and Highly Stable Blue Light-Emitting Diodes}. \emph{ACS Energy Lett.}
  \textbf{2020}, \emph{5}, 1062\relax
\mciteBstWouldAddEndPuncttrue
\mciteSetBstMidEndSepPunct{\mcitedefaultmidpunct}
{\mcitedefaultendpunct}{\mcitedefaultseppunct}\relax
\EndOfBibitem
\bibitem[Xu \latin{et~al.}(2021)Xu, Xi, Dong, Ahn, Zhu, Chen, Li, Zhu, Dai, Hu,
  Jiao, Hou, Li, and Wu]{XuJ2021}
Xu,~J.; Xi,~J.; Dong,~H.; Ahn,~N.; Zhu,~Z.; Chen,~J.; Li,~P.; Zhu,~X.; Dai,~J.;
  Hu,~Z.; Jiao,~B.; Hou,~X.; Li,~J.; Wu,~Z. {Impermeable inorganic ``Walls''
  Sandwiching Perovskite Layer Toward Inverted and Indoor Photovoltaic
  Devices}. \emph{Nano Energy} \textbf{2021}, \emph{88}, 106286\relax
\mciteBstWouldAddEndPuncttrue
\mciteSetBstMidEndSepPunct{\mcitedefaultmidpunct}
{\mcitedefaultendpunct}{\mcitedefaultseppunct}\relax
\EndOfBibitem
\bibitem[Seidu \latin{et~al.}(2021)Seidu, Dvorak, Rinke, and Li]{Seidu2021}
Seidu,~A.; Dvorak,~M.; Rinke,~P.; Li,~J. {Atomic and Electronic Structure of
  Cesium Lead Triiodide Surfaces}. \emph{J. Chem. Phys.} \textbf{2021},
  \emph{154}, 074712\relax
\mciteBstWouldAddEndPuncttrue
\mciteSetBstMidEndSepPunct{\mcitedefaultmidpunct}
{\mcitedefaultendpunct}{\mcitedefaultseppunct}\relax
\EndOfBibitem
\bibitem[Seidu \latin{et~al.}(2021)Seidu, Dvorak, J\"{a}rvi, Rinke, and
  Li]{ASeidu2021}
Seidu,~A.; Dvorak,~M.; J\"{a}rvi,~J.; Rinke,~P.; Li,~J. {Surface Reconstruction
  of Tetragonal Methylammonium Lead Triiodide}. \emph{APL Mater.}
  \textbf{2021}, \emph{9}, 111102\relax
\mciteBstWouldAddEndPuncttrue
\mciteSetBstMidEndSepPunct{\mcitedefaultmidpunct}
{\mcitedefaultendpunct}{\mcitedefaultseppunct}\relax
\EndOfBibitem
\bibitem[Todorovi\'{c} \latin{et~al.}(2019)Todorovi\'{c}, Gutmann, Corander,
  and Rinke]{Todorovic2019}
Todorovi\'{c},~M.; Gutmann,~M.~U.; Corander,~J.; Rinke,~P. {Bayesian Inference
  of Atomistic Structure in Functional Materials}. \emph{{npj Comput. Mater.}}
  \textbf{2019}, \emph{35}, 1\relax
\mciteBstWouldAddEndPuncttrue
\mciteSetBstMidEndSepPunct{\mcitedefaultmidpunct}
{\mcitedefaultendpunct}{\mcitedefaultseppunct}\relax
\EndOfBibitem
\bibitem[Fang \latin{et~al.}(2021)Fang, Makkonen, Todorov\'c, Rinke, and
  Chen]{FangL2021}
Fang,~L.; Makkonen,~E.; Todorov\'c,~M.; Rinke,~P.; Chen,~X. Efficient Amino
  Acid Conformer Search with Bayesian Optimization. \emph{J. Chem. Theo.
  Comput.} \textbf{2021}, \emph{17}, 1955\relax
\mciteBstWouldAddEndPuncttrue
\mciteSetBstMidEndSepPunct{\mcitedefaultmidpunct}
{\mcitedefaultendpunct}{\mcitedefaultseppunct}\relax
\EndOfBibitem
\bibitem[Egger \latin{et~al.}(2020)Egger, H\"ormann, Jeindl, Scherbela,
  Obersteiner, Todorovi\'{c}, Rinke, and Hofmann]{EggerAT2020}
Egger,~A.~T.; H\"ormann,~L.; Jeindl,~A.; Scherbela,~M.; Obersteiner,~V.;
  Todorovi\'{c},~M.; Rinke,~P.; Hofmann,~O.~T. {Charge Transfer into Organic
  Thin Films: A Deeper Insight through Machine-Learning-Assisted Structure
  Search}. \emph{Adv. Sci.} \textbf{2020}, \emph{7}, 2000992\relax
\mciteBstWouldAddEndPuncttrue
\mciteSetBstMidEndSepPunct{\mcitedefaultmidpunct}
{\mcitedefaultendpunct}{\mcitedefaultseppunct}\relax
\EndOfBibitem
\bibitem[J\"arvi \latin{et~al.}(2020)J\"arvi, Rinke, and
  Todorovi\'c]{Jarvi2020}
J\"arvi,~J.; Rinke,~P.; Todorovi\'c,~M. {Detecting Stable Adsorbates of
  (1S)-camphor on $\text{Cu}(111)$ with Bayesian Optimization}.
  \emph{{Beilstein J. Nanotechnol.}} \textbf{2020}, \emph{11}, 1577\relax
\mciteBstWouldAddEndPuncttrue
\mciteSetBstMidEndSepPunct{\mcitedefaultmidpunct}
{\mcitedefaultendpunct}{\mcitedefaultseppunct}\relax
\EndOfBibitem
\bibitem[J\"arvi \latin{et~al.}(2020)J\"arvi, Alldritt, Krej\v{c}\'i,
  Todorovi\'c, Liljeroth, and Rinke]{Jarvi2021}
J\"arvi,~J.; Alldritt,~B.; Krej\v{c}\'i,~O.; Todorovi\'c,~M.; Liljeroth,~P.;
  Rinke,~P. {Integrating Bayesian Inference with Scanning Probe Experiments for
  Robust Identification of Surface Adsorbate Configurations}. \emph{Adv. Funct.
  Mater.} \textbf{2020}, \emph{31}, 2010853\relax
\mciteBstWouldAddEndPuncttrue
\mciteSetBstMidEndSepPunct{\mcitedefaultmidpunct}
{\mcitedefaultendpunct}{\mcitedefaultseppunct}\relax
\EndOfBibitem
\bibitem[Sun \latin{et~al.}(2011)Sun, Seo, Takacs, Seifter, and
  Heeger]{Sun2011}
Sun,~Y.; Seo,~J.~H.; Takacs,~C.~J.; Seifter,~J.; Heeger,~A.~J. {Inverted
  Polymer Solar Cells Integrated with a Low-Temperature-Annealed
  Sol-Gel-Derived ZnO Film as an Electron Transport Layer}. \emph{Adv. Mater.}
  \textbf{2011}, \emph{23}, 1679--1683\relax
\mciteBstWouldAddEndPuncttrue
\mciteSetBstMidEndSepPunct{\mcitedefaultmidpunct}
{\mcitedefaultendpunct}{\mcitedefaultseppunct}\relax
\EndOfBibitem
\bibitem[Ro \latin{et~al.}(2016)Ro, Downing, Engmann, Herzing, DeLongchamp,
  Richter, Mukherjee, Ade, Abdelsamie, Jagadamma, Amassian, Liue, and
  Yan]{Ro2016}
Ro,~H.~W.; Downing,~J.~M.; Engmann,~S.; Herzing,~A.~A.; DeLongchamp,~D.~M.;
  Richter,~L.~J.; Mukherjee,~S.; Ade,~H.; Abdelsamie,~M.; Jagadamma,~L.~K.;
  Amassian,~A.; Liue,~Y.; Yan,~H. {Morphology Changes upon Scaling a
  High-Efficiency, Solution-Processed Solar Cell}. \emph{Energy Environ. Sci.}
  \textbf{2016}, \emph{9}, 2835--2846\relax
\mciteBstWouldAddEndPuncttrue
\mciteSetBstMidEndSepPunct{\mcitedefaultmidpunct}
{\mcitedefaultendpunct}{\mcitedefaultseppunct}\relax
\EndOfBibitem
\bibitem[Liang \latin{et~al.}(2015)Liang, Zhang, Jiang, and Cao]{Liang2015}
Liang,~Z.; Zhang,~Q.; Jiang,~L.; Cao,~G. {ZnO Cathode Buffer Layers for
  Inverted Polymer Solar Cells}. \emph{Energy Environ. Sci.} \textbf{2015},
  \emph{8}, 3442\relax
\mciteBstWouldAddEndPuncttrue
\mciteSetBstMidEndSepPunct{\mcitedefaultmidpunct}
{\mcitedefaultendpunct}{\mcitedefaultseppunct}\relax
\EndOfBibitem
\bibitem[Yu \latin{et~al.}(2015)Yu, Huang, Yang, Fu, Zhou, Zhang, and
  Li]{Yu2015}
Yu,~W.; Huang,~L.; Yang,~D.; Fu,~P.; Zhou,~L.; Zhang,~J.; Li,~C. {Efficiency
  Exceeding 10\% for Inverted Polymer Solar Cells with a ZnO/Ionic Liquid
  Combined Cathode Interfacial Layer}. \emph{Energy Environ. Sci.}
  \textbf{2015}, \emph{3}, 10660\relax
\mciteBstWouldAddEndPuncttrue
\mciteSetBstMidEndSepPunct{\mcitedefaultmidpunct}
{\mcitedefaultendpunct}{\mcitedefaultseppunct}\relax
\EndOfBibitem
\bibitem[Zheng \latin{et~al.}(2019)Zheng, Wang, Huang, Wang, Ke, Logsdon, Wang,
  Wang, Zhu, Yu, Wasielewski, Kanatzidis, Marks, and Facchett]{Zheng2019}
Zheng,~D.; Wang,~G.; Huang,~W.; Wang,~B.; Ke,~W.; Logsdon,~J.~L.; Wang,~H.;
  Wang,~Z.; Zhu,~W.; Yu,~J.; Wasielewski,~M.~R.; Kanatzidis,~M.~G.;
  Marks,~T.~J.; Facchett,~A. {Combustion Synthesized Zinc Oxide
  Electron-Transport Layers for Efﬁcient and Stable Perovskite Solar Cells}.
  \emph{Adv. Funct. Mater.} \textbf{2019}, \emph{29}, 1900265\relax
\mciteBstWouldAddEndPuncttrue
\mciteSetBstMidEndSepPunct{\mcitedefaultmidpunct}
{\mcitedefaultendpunct}{\mcitedefaultseppunct}\relax
\EndOfBibitem
\bibitem[{Mej\'ia~Escobar} \latin{et~al.}(2017){Mej\'ia~Escobar}, Pathak, Liu,
  Snaith, and Jaramillo]{Escobar17}
{Mej\'ia~Escobar},~M.~A.; Pathak,~S.; Liu,~J.; Snaith,~H.~J.; Jaramillo,~F.
  {ZrO$_2$/TiO$_2$ Electron Collection Layer for Efficient Meso-Superstructured
  Hybrid Perovskite Solar Cells}. \emph{ACS Appl. Mater. Interfaces}
  \textbf{2017}, \emph{9}, 2342\relax
\mciteBstWouldAddEndPuncttrue
\mciteSetBstMidEndSepPunct{\mcitedefaultmidpunct}
{\mcitedefaultendpunct}{\mcitedefaultseppunct}\relax
\EndOfBibitem
\bibitem[Li \latin{et~al.}(2018)Li, Zhao, Wei, Xiao, Dong, Wan, and
  Wang]{LiY2018}
Li,~Y.; Zhao,~L.; Wei,~S.; Xiao,~M.; Dong,~B.; Wan,~L.; Wang,~S. {Effect of
  ZrO$_2$ Film Thickness on the Photoelectric Properties of Mixed-Cation
  Perovskite Solar Cells}. \emph{Appl. Surf. Sci.} \textbf{2018}, \emph{439},
  506--515\relax
\mciteBstWouldAddEndPuncttrue
\mciteSetBstMidEndSepPunct{\mcitedefaultmidpunct}
{\mcitedefaultendpunct}{\mcitedefaultseppunct}\relax
\EndOfBibitem
\bibitem[Navrotsky(2004)]{Narotsky2004}
Navrotsky,~A. {Energetic Clues to Pathways to Biomineralization: Precursors,
  Clusters, and Nanoparticles}. \emph{Proc. Natl. Acad. Sci.} \textbf{2004},
  \emph{101}, 12096\relax
\mciteBstWouldAddEndPuncttrue
\mciteSetBstMidEndSepPunct{\mcitedefaultmidpunct}
{\mcitedefaultendpunct}{\mcitedefaultseppunct}\relax
\EndOfBibitem
\bibitem[Demiroglu and Bromley(2013)Demiroglu, and Bromley]{Demiroglu2013}
Demiroglu,~I.; Bromley,~S.~T. {Nanofilm versus Bulk Polymorphism in Wurtzite
  Materials}. \emph{Phys. Rev. Lett.} \textbf{2013}, \emph{110}, 245501\relax
\mciteBstWouldAddEndPuncttrue
\mciteSetBstMidEndSepPunct{\mcitedefaultmidpunct}
{\mcitedefaultendpunct}{\mcitedefaultseppunct}\relax
\EndOfBibitem
\bibitem[Wu \latin{et~al.}(2011)Wu, Lagally, and Liu]{Wu2011}
Wu,~D.; Lagally,~M.~G.; Liu,~F. {Stabilizing Graphitic Thin Films of Wurtzite
  Materials by Epitaxial Strain}. \emph{Phys. Rev. Lett.} \textbf{2011},
  \emph{107}, 236101\relax
\mciteBstWouldAddEndPuncttrue
\mciteSetBstMidEndSepPunct{\mcitedefaultmidpunct}
{\mcitedefaultendpunct}{\mcitedefaultseppunct}\relax
\EndOfBibitem
\bibitem[Freeman \latin{et~al.}(2006)Freeman, Claeyssens, and
  Harding]{Freeman2006}
Freeman,~C.~L.; Claeyssens,~F.; Harding,~N. L.~A. J.~H. {Graphitic Nanofilms as
  Precursors to Wurtzite Films: Theory}. \emph{Phys. Rev. Lett.} \textbf{2006},
  \emph{96}, 066102\relax
\mciteBstWouldAddEndPuncttrue
\mciteSetBstMidEndSepPunct{\mcitedefaultmidpunct}
{\mcitedefaultendpunct}{\mcitedefaultseppunct}\relax
\EndOfBibitem
\bibitem[Bieniek \latin{et~al.}(2015)Bieniek, Hofmann, and Rinke]{Bieniek2015}
Bieniek,~B.; Hofmann,~O.~T.; Rinke,~P. {Influence of Hydrogen on the Structure
  and Stability of Ultra-Thin ZnO on Metal Substrates}. \emph{Appl. Phys.
  Lett.} \textbf{2015}, \emph{106}, 131602\relax
\mciteBstWouldAddEndPuncttrue
\mciteSetBstMidEndSepPunct{\mcitedefaultmidpunct}
{\mcitedefaultendpunct}{\mcitedefaultseppunct}\relax
\EndOfBibitem
\bibitem[Heyd \latin{et~al.}(2003)Heyd, Scuseria, and Ernzerhof]{Heyd2003}
Heyd,~J.; Scuseria,~G.~E.; Ernzerhof,~M. {Hybrid Functionals Based on a
  Screened Coulomb Potential}. \emph{J. Chem. Phys.} \textbf{2003}, \emph{118},
  8207\relax
\mciteBstWouldAddEndPuncttrue
\mciteSetBstMidEndSepPunct{\mcitedefaultmidpunct}
{\mcitedefaultendpunct}{\mcitedefaultseppunct}\relax
\EndOfBibitem
\bibitem[Perdew \latin{et~al.}(2008)Perdew, Ruzsinszky, Csonka, Vydrov,
  Scuseria, Constantin, Zhou, and Burke]{perdew08}
Perdew,~J.~P.; Ruzsinszky,~A.; Csonka,~G.~I.; Vydrov,~O.~A.; Scuseria,~G.~E.;
  Constantin,~L.~A.; Zhou,~X.; Burke,~K. {Restoring the Density-Gradient
  Expansion for Exchange in Solids and Surfaces}. \emph{Phys. Rev. Lett.}
  \textbf{2008}, \emph{100}, 136406\relax
\mciteBstWouldAddEndPuncttrue
\mciteSetBstMidEndSepPunct{\mcitedefaultmidpunct}
{\mcitedefaultendpunct}{\mcitedefaultseppunct}\relax
\EndOfBibitem
\bibitem[Levchenko \latin{et~al.}(2015)Levchenko, Ren, Wieferink, Johanni,
  Rinke, Blum, and Scheffler]{Levchenko/etal:2015}
Levchenko,~S.~V.; Ren,~X.; Wieferink,~J.; Johanni,~R.; Rinke,~P.; Blum,~V.;
  Scheffler,~M. {Hybrid Functionals for Large Periodic Systems in an
  All-electron, Numeric Atom-centered Basis Framework }. \emph{Comput. Phys.
  Commun.} \textbf{2015}, \emph{192}, 60 -- 69\relax
\mciteBstWouldAddEndPuncttrue
\mciteSetBstMidEndSepPunct{\mcitedefaultmidpunct}
{\mcitedefaultendpunct}{\mcitedefaultseppunct}\relax
\EndOfBibitem
\bibitem[Neugebauer and Scheffler(1992)Neugebauer, and Scheffler]{Neugebauer92}
Neugebauer,~J.; Scheffler,~M. {Adsorbate-substrate and Adsorbate-adsorbate
  Interactions of Na and K Adlayers on Al(111)}. \emph{Phys. Rev. B}
  \textbf{1992}, \emph{46}, 16067\relax
\mciteBstWouldAddEndPuncttrue
\mciteSetBstMidEndSepPunct{\mcitedefaultmidpunct}
{\mcitedefaultendpunct}{\mcitedefaultseppunct}\relax
\EndOfBibitem
\bibitem[Marronnier \latin{et~al.}(2018)Marronnier, Roma, Boyer-Richard,
  Pedesseau, Jancu, Bonnassieux, Katan, Stoumpos, Kanatzidis, and
  Even]{Marronnier2018}
Marronnier,~A.; Roma,~G.; Boyer-Richard,~S.; Pedesseau,~L.; Jancu,~J.-M.;
  Bonnassieux,~Y.; Katan,~C.; Stoumpos,~C.~C.; Kanatzidis,~M.~G.; Even,~J.
  {Anharmonicity and Disorder in the Black Phases of Cesium Lead Iodide Used
  for Stable Inorganic Perovskite Solar Cells}. \emph{{ACS Nano}}
  \textbf{2018}, \emph{12}, 3477--3486\relax
\mciteBstWouldAddEndPuncttrue
\mciteSetBstMidEndSepPunct{\mcitedefaultmidpunct}
{\mcitedefaultendpunct}{\mcitedefaultseppunct}\relax
\EndOfBibitem
\bibitem[Not()]{Note-NOMAD}
Data to be uploaded on Nomad\relax
\mciteBstWouldAddEndPuncttrue
\mciteSetBstMidEndSepPunct{\mcitedefaultmidpunct}
{\mcitedefaultendpunct}{\mcitedefaultseppunct}\relax
\EndOfBibitem
\end{mcitethebibliography}
\end{document}





\clearpage

\section{Binding energy landscapes from BOSS-DFT structural search}
\begin{figure}[!ht]
\centering
\includegraphics[width=\linewidth]{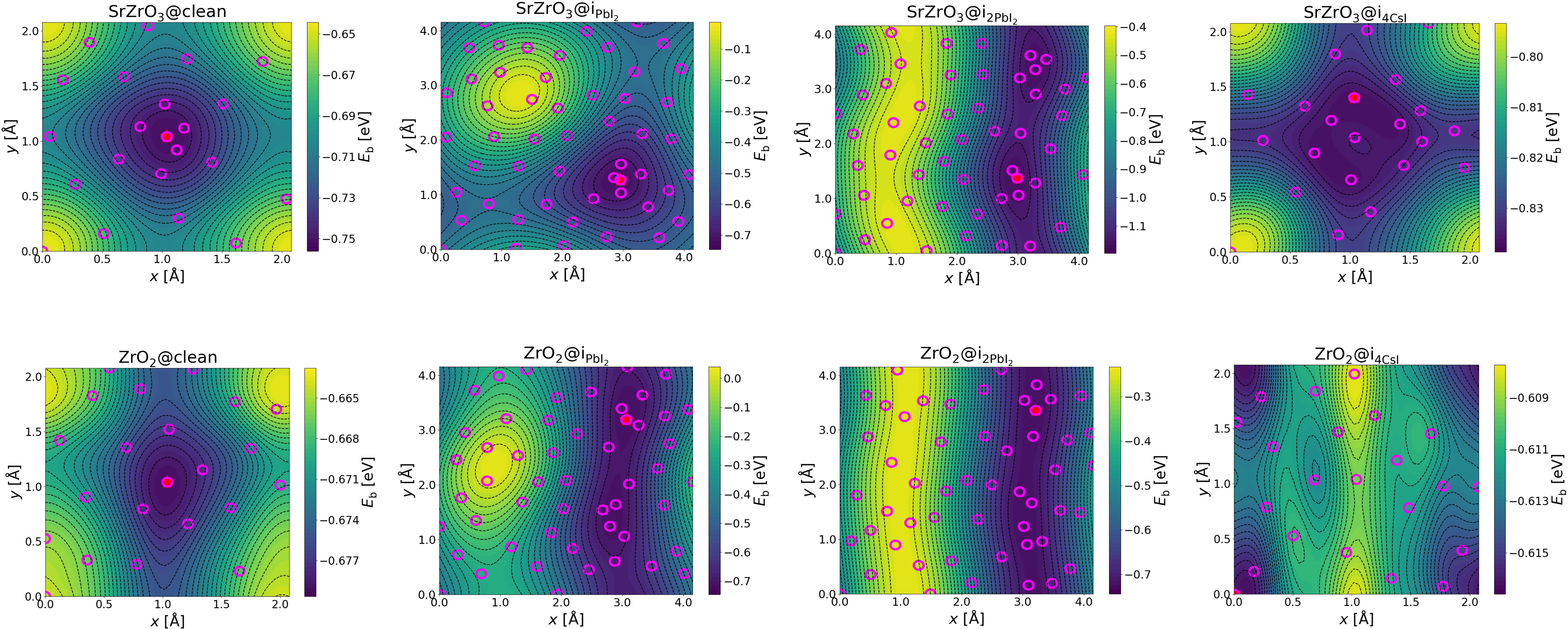}\vspace{0.1em}
\caption {Energy landscapes of perovskite-coating interfaces with BOSS and DFT. The pink circles and red star depict the acquisition points and minimum potential energies at $x$ and $y$ respectively. The top panel consist of SrZrO$_3$ on CsI-T, $i_{\text{PbI}_2^{}}$, $i_{2\text{PbI}_2^{}}$ and $i_{4\text{CsI}}$. The bottom panel comprises of ZrO$_2$ on CsI-T, $i_{\text{PbI}_2^{}}$, $i_{2\text{PbI}_2^{}}$ and $i_{4\text{CsI}}$.} 
\label{landscape-1}
\end{figure}

Figure~\ref{landscape-1} shows the energy landscapes for SrZrO$_3$ and ZrO$_2$ on the $\upalpha$-CsPbI$_3$ most stable reconstructed surface models (clean, $i_{\text{PbI}_2^{}}$, $i_{2\text{PbI}_2^{}}$ and $i_{4\text{CsI}}$). $x$, $y$ and $E_b$ depict the varied x-, y-coordinates and the potential energy at a given $x$, $y$ pair, respectively. The pink circles and red star are representative of the acquisition points and the optimal potential energy at $x$ and $y$ respectively.

Table~\ref{boss_energy} lists the minima of the surrogate models from the BOSS/DFT search (i.e. positions of the red stars of Fig.~\ref{landscape-1}).

\begin{table}[!htp]
\caption{Minimum binding energies (in eV) and their corresponding locations (acquisition points) $(x,y)$ of the surrogate model from the BOSS/DFT structural search.} \label{boss_energy}
\begin{tabular}{c|ccc|ccc|ccc} \hline\hline
& \multicolumn{3}{c|}{SrZrO$_3$} & \multicolumn{3}{c|}{ZrO$_2$} & \multicolumn{3}{c}{ZnO} \\ \hline 
& \multicolumn{2}{c}{Location [{\AA}]} & $E_{\text{b}}$ & \multicolumn{2}{c}{Location [{\AA}]} & $E_{\text{b}}$  & \multicolumn{2}{c}{Location [{\AA}]} & $E_{\text{b}}$ \\
& $x$ & $y$ & [eV] & $x$ & $y$ & [eV] & $x$ & $y$ & [eV]\\ \hline
clean & ~1.04 &~1.04 &~$-0.75$ &~1.04 & ~1.04& ~$-0.68 $ &~1.04 & ~1.04 &~$-0.56$\\
i$_{\text{PbI}_2^{}}$ & ~2.96 & ~1.27 &~$-0.72$ & ~3.09 &~3.19 &~$-0.72$ &~2.18 & ~1.65 &~$-0.34$ \\
i$_{2\text{PbI}_2^{}}$ & ~2.99 & ~1.37 &~$-1.17$ & ~3.21 & ~3.36 &~$-0.72$ & ~2.06 & ~1.63 &~$-0.69 $\\
i$_{4\text{CsI}}$ & ~3.08 & ~2.78 & ~$-0.84$ & ~3.24 & ~4.13 &~$-2.43$ & ~0.76 & ~3.17 &~$-0.67$\\\hline\hline
\end{tabular}
\end{table}

\clearpage

\section{Optimized interface structures}
\begin{figure}[!ht]
\centering
\includegraphics[width=0.6\linewidth]{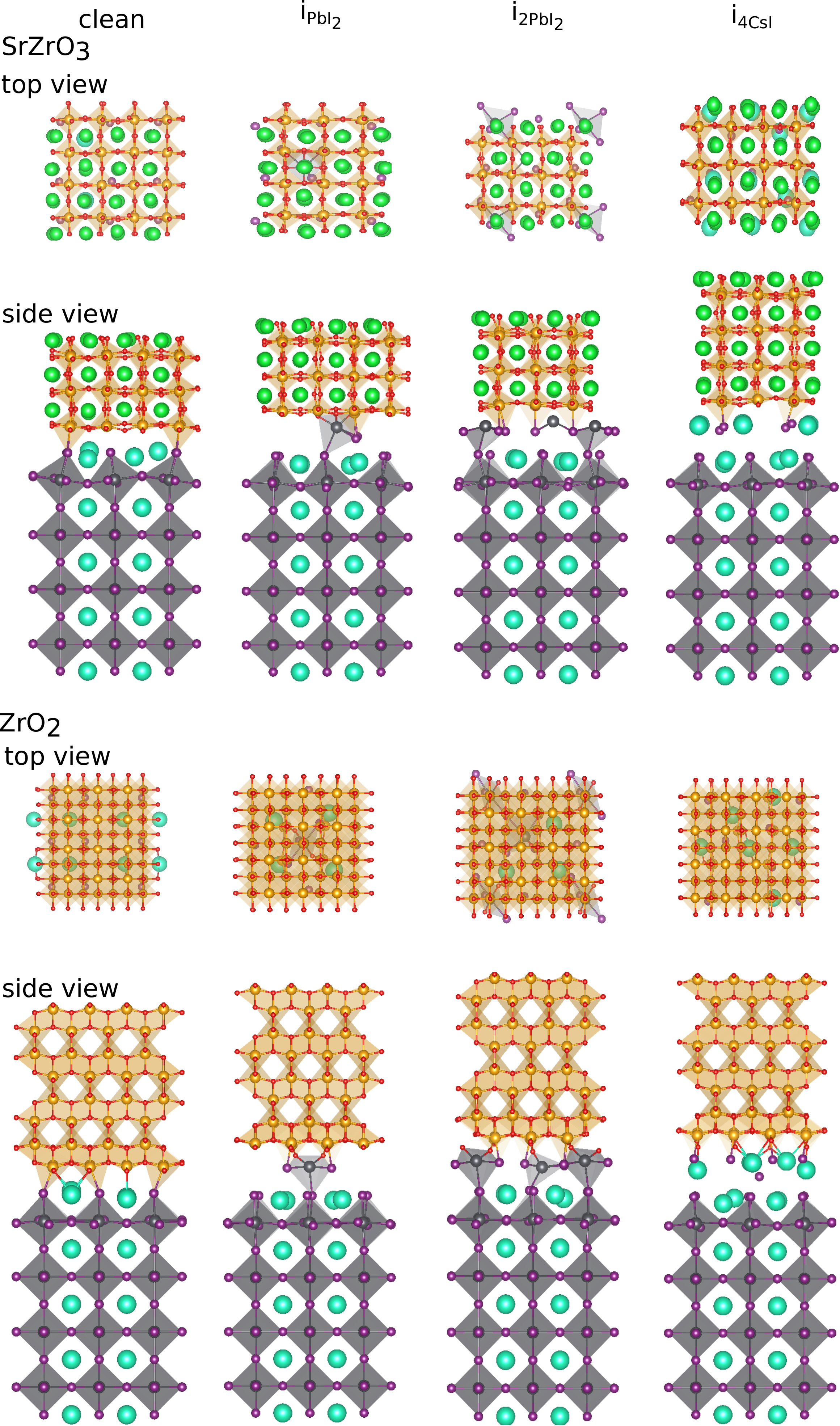}\vspace{0.1em}
\caption {Optimized coating/perovskite interfaces for SrZrO$_3$ (top panel) and ZrO$_2$ (bottom panel on CsI-T, $i_{\text{PbI}_2^{}}$, $i_{2\text{PbI}_2^{}}$ and $i_{4\text{CsI}}$. } 
\label{structure-1}
\end{figure}

Figure~\ref{structure-1} depicts the optimized atomic structures of SrZrO$_3$ (top panel) and ZrO$_2$ (bottom panel) on CsI-T, $i_{\text{PbI}_2^{}}$, $i_{2\text{PbI}_2^{}}$ and $i_{4\text{CsI}}$,  respectively.

\clearpage
\section{Binding energies of optimized structures vs lattice strain}
\begin{figure}[!ht]
\centering
\includegraphics[width=0.6\linewidth]{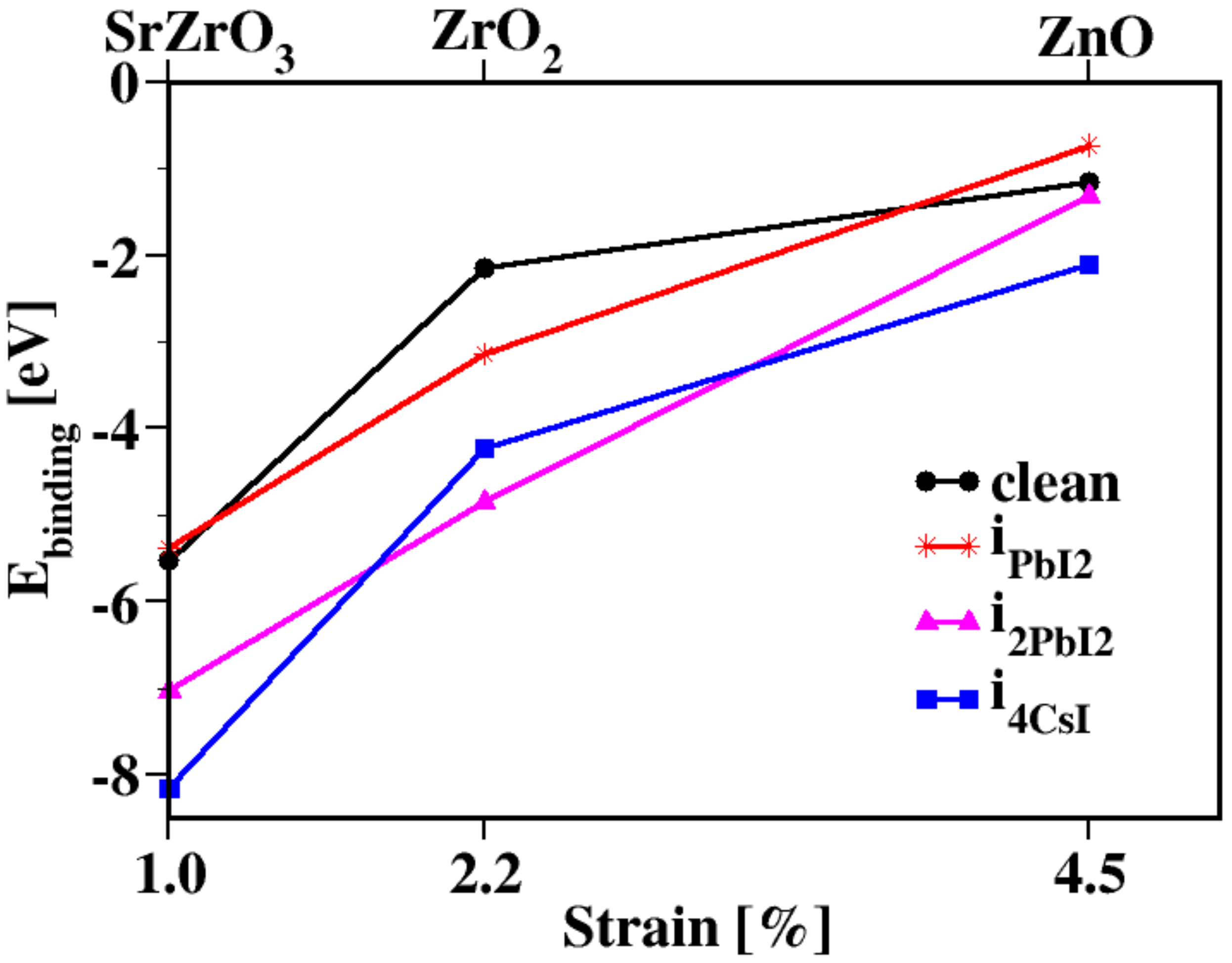}\vspace{0.1em}
\caption {Binding energies of optimized structures as a function of lattice strain. Coatings on clean surface,  $i_{\text{PbI}_2^{}}$, $i_{2\text{PbI}_2^{}}$ and $i_{4\text{CsI}}$ are depicted by black, red, blue and pink colours respectively. The x-axis shows the absolute value of the lattice strain of each coating and the y-axis the binding energy for that substrate-coating pair. }
\label{Ebind_1}
\end{figure}

Figure~\ref{Ebind_1} shows the binding energies of the optimized structures as a function of lattice strain. Coatings on the clean surface,  $i_{\text{PbI}_2^{}}$, $i_{2\text{PbI}_2^{}}$ and $i_{4\text{CsI}}$ are depicted by black, red, blue and pink colours, respectively. The x-axis shows the absolute value of the lattice mismatch between the substrate and each coating.
 \newpage
\section{Estimation of the hybrid exchange correlation coefficient}
\begin{figure}[!ht]
\centering
\includegraphics[width=0.8\linewidth]{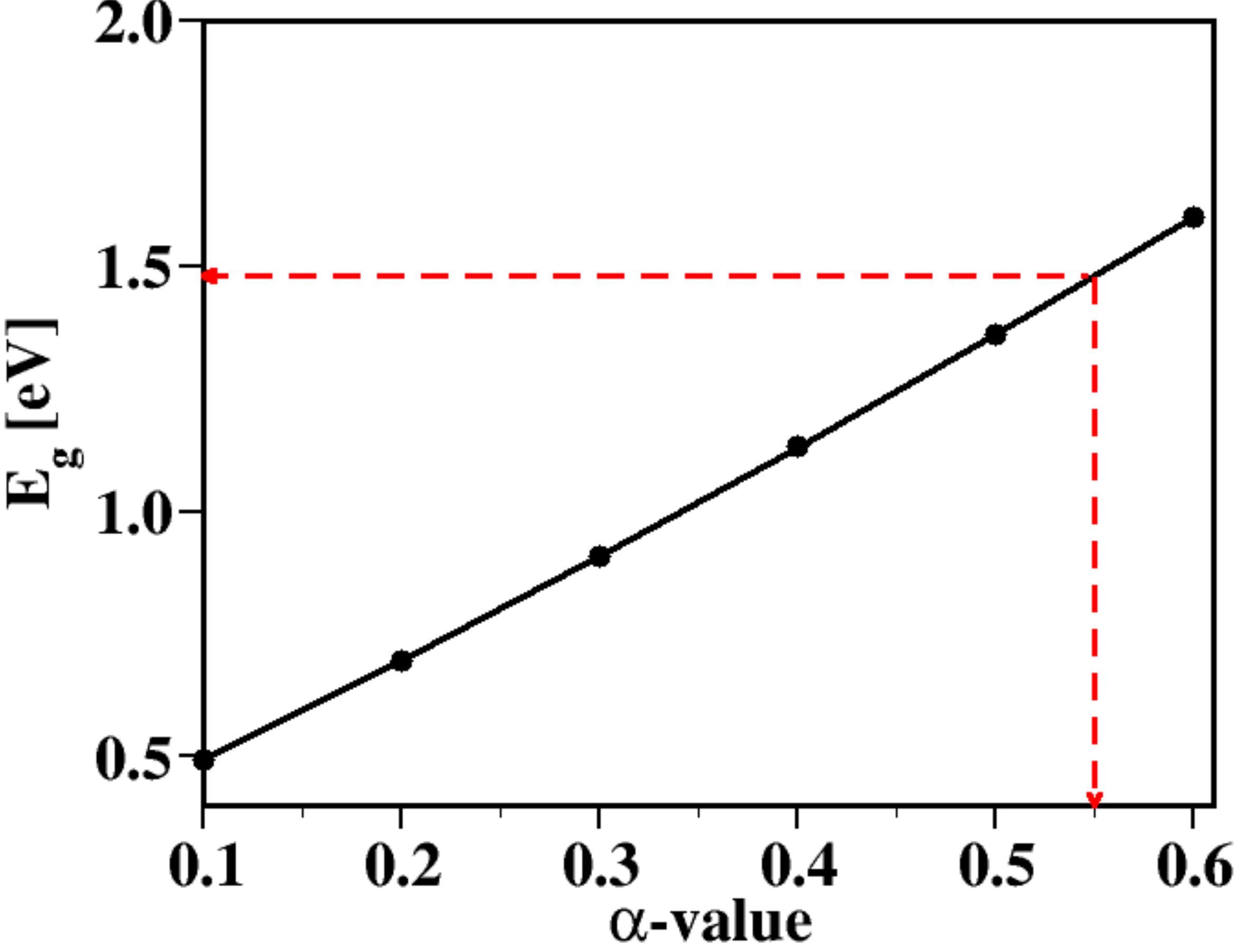}\vspace{0.1em}
\caption {Estimated hybrid exchange correlation coefficient ($\alpha$ value for HSE band structure calculation. The dash red line depicts the estimated $\alpha$ value (on the horizontal axis) and its corresponding band gap energy (on the vertical axis).} \label{alpha}
\end{figure}

We estimated the hybrid exchange correlation coefficient ($\alpha$ value) by fitting our HSE+SOC band gap to the $GW$ band gap of Ref.~\textcolor{blue}{74}. This we achieved by using the bulk cubic unit cell provided in Ref.~\textcolor{blue}{74}. The lattice constants of the Reference bulk unit cell are $a=b=c= 6.30$ {\AA}. We varied the $\alpha$ value from $0.1-0.6$ as shown in Fig.~\ref{alpha}. The dash red line depicts the estimated $\alpha$ value of 0.55 which corresponds to a band gap energy of $\sim1.48$ eV.

In our work, we reported an HSE+SOC bulk band gap energy of $\sim 1.34$ eV, which differs from the Reference value by $\sim0.14$ eV. This difference can be attributed to the difference in the lattice constants of the Reference structure and our PBEsol optimized bulk structure ($a=b=c=6.22$ {\AA}).

\clearpage
\section{Band structures of bulk and all interfaces}
\begin{figure}[!ht]
\centering
\includegraphics[width=\textwidth]{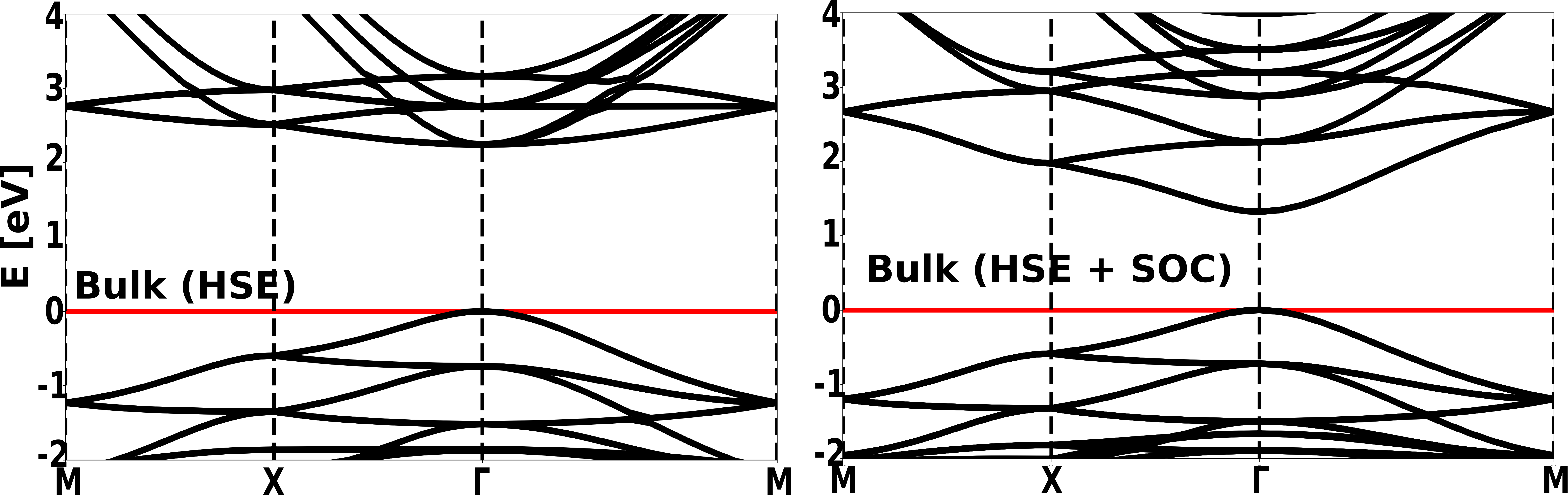}\vspace{0.1em}
\caption {Bulk band structure with and without spin-orbit coupling (SOC). The red horizontal line is representative of the valence band maximum.} \label{bulk}
\end{figure}

Figure~\ref{bulk} shows the bulk band structure calculated with (right panel) and without (left panel) spin-orbit coupling (SOC). The red horizontal line depicts the valence band maximum (VBM). Upon inclusion of SOC, the conduction band minimum (CBM) is shifted down by $\sim 0.8$ eV while the VBM shifts up by $\sim 0.1$ eV, reducing the bulk band gap energy to 1.34 eV. 

\begin{figure}[!ht]
\includegraphics[width=\textwidth]{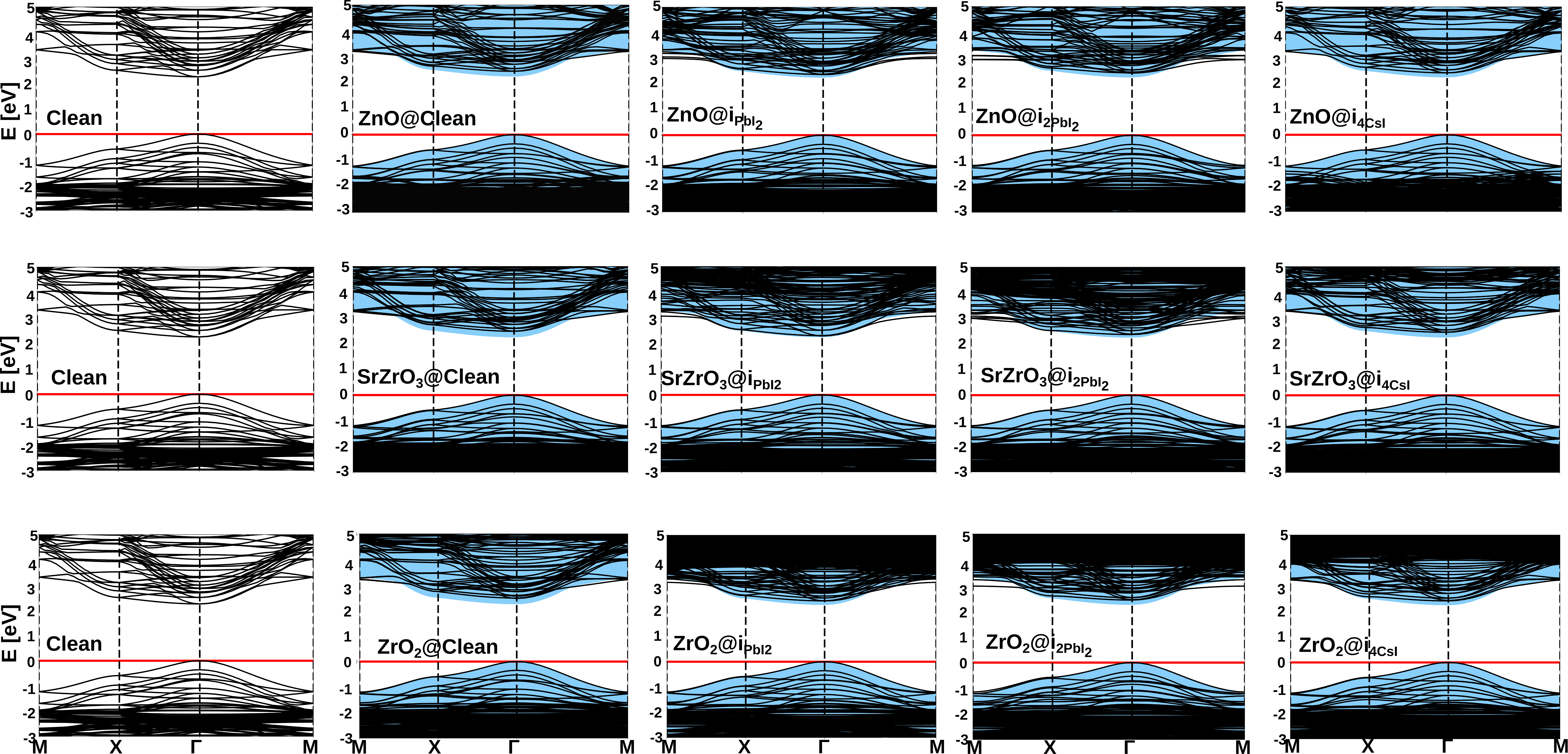}\vspace{0.1em}
\caption {Band structures for ZnO, SrZrO$_3$ and ZrO$_2$ on the $\upalpha$-CsPbI$_3$ reconstructed surface (CsI-T, $i_{\text{PbI}_2^{}}$, $i_{2\text{PbI}_2^{}}$ and $i_{4\text{CsI}}$) models. The red horizontal depicts the valence band maximum (VBM) The band structures of the coating-perovskite interface are overlayed on to the clean surface model for easy of comparison.} \label{bands}
\end{figure}

Figure~\ref{bands} shows the band structures of ZnO, SrZrO$_3$ and ZrO$_2$ on the $\upalpha$-CsPbI$_3$ reconstructed surface (CsI-T, $i_{\text{PbI}_2^{}}$, $i_{2\text{PbI}_2^{}}$ and $i_{4\text{CsI}}$) models. The red horizontal depicts the valence band maximum (VBM). The band structures of the coating-perovskite interface are overlayed on to the clean surface model for ease of comparison.

\clearpage
\section{Spatially resolved local density of states and band alignment}

\begin{figure}[!ht]
\includegraphics[width=\linewidth]{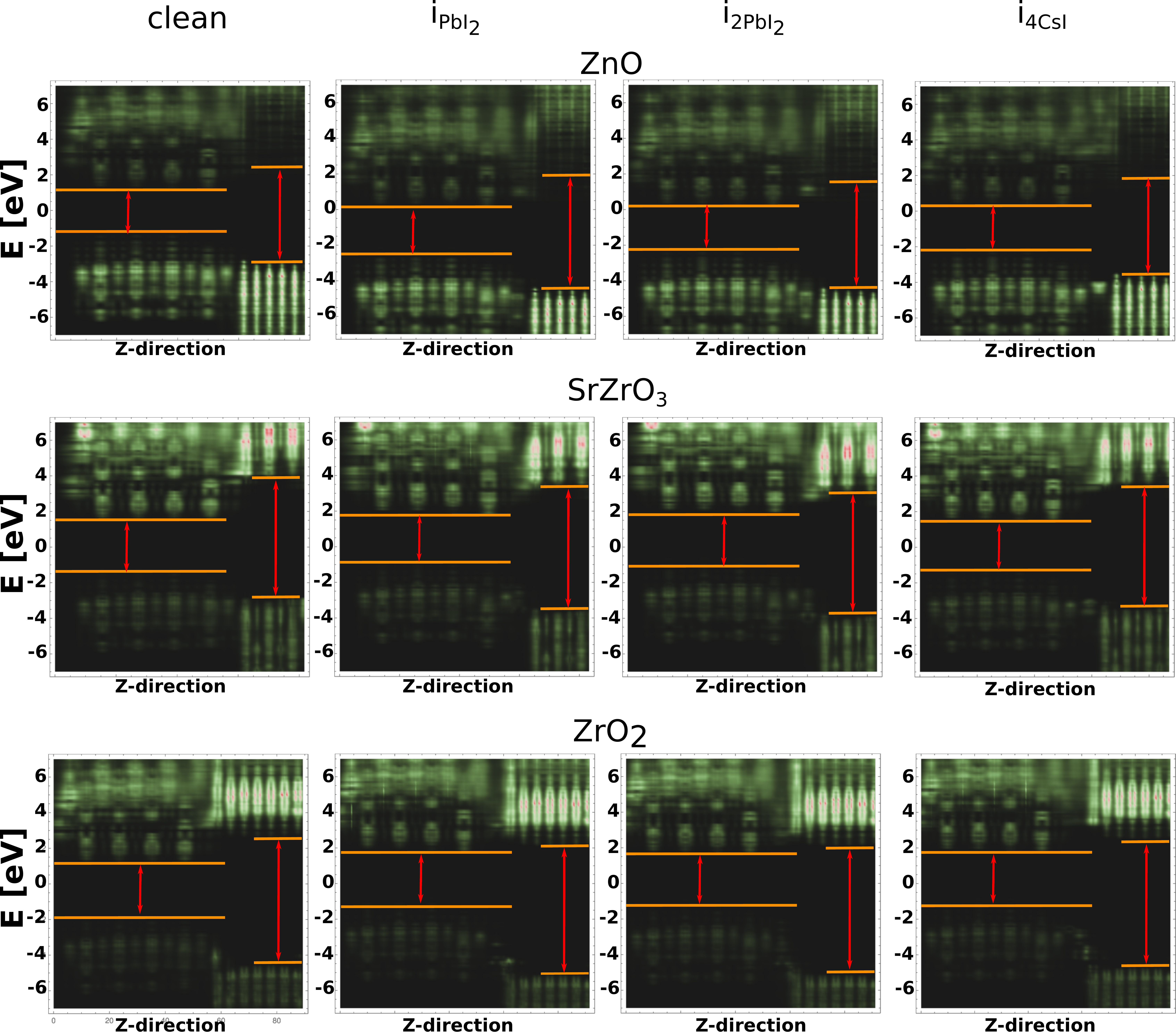}\vspace{0.1em}
\caption {Spatially resolved local density of states of all coatings on the $\upalpha$-CsPbI$_3$ reconstructed surface (CsI-T, $i_{\text{PbI}_2^{}}$, $i_{2\text{PbI}_2^{}}$ and $i_{4\text{CsI}}$) models. The horizontal orange lines depict VBM and CBM for the coatings and perovskites. The red vertical lines are representative of the energy band gap of the reconstructed $\upalpha$-CsPbI$_3$ surface models and coatings.} \label{ldos_1}
\end{figure}

Figure~\ref{ldos_1} depicts the spatially resolved local density of states (LDOS) of all coatings on the $\upalpha$-CsPbI$_3$ reconstructed surface (CsI-T, $i_{\text{PbI}_2^{}}$, $i_{2\text{PbI}_2^{}}$ and $i_{4\text{CsI}}$) models. The horizontal orange lines depict the level alignments of the VBM and CBM of coatings and perovskites. The vertical red lines are representative of the band gap of the perovskites and coatings. Due to the enormous computational cost of the HSE functional, these interface calculations were performed without SOC. In the main text of the manuscript, we explain how we add the SOC for CsPbI$_3$ to the band offsets.

